\documentclass[fleqn,usenatbib]{mnras}

\usepackage{newtxtext,newtxmath}
\usepackage[T1]{fontenc}

\DeclareRobustCommand{\VAN}[3]{#2}
\let\VANthebibliography\thebibliography
\def\thebibliography{\DeclareRobustCommand{\VAN}[3]{##3}\VANthebibliography}

\usepackage{graphicx}	% Including figure files
\usepackage{amsmath}	% Advanced maths commands
\title[M-dwarfs measured with \textit{CHEOPS}]{The EBLM project -- IX. Five fully convective M-dwarfs, precisely measured with \textit{CHEOPS} and \textit{TESS} light curves\thanks{Based on observations collected at the Observatoire de Haute-Provence (CNRS, France) and at the Southern African Large Telescope (SALT)}}

\author[D. Sebastian et al.]{
D. Sebastian$^{1}$ $^{\href{https://orcid.org/0000-0002-2214-9258}{\includegraphics[scale=0.5]{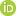}}}$\thanks{E-mail: D.Sebastian.1@bham.ac.uk}, 
M. I. Swayne$^{2}$ $^{\href{https://orcid.org/0000-0002-2609-3159}{\includegraphics[scale=0.5]{Images/orcid.jpg}}}$,
P. F. L. Maxted$^{2}$ $^{\href{https://orcid.org/0000-0003-3794-1317}{\includegraphics[scale=0.5]{Images/orcid.jpg}}}$,
A. H. M. J. Triaud$^{1}$ $^{\href{https://orcid.org/0000-0002-5510-8751}{\includegraphics[scale=0.5]{Images/orcid.jpg}}}$,
S. G. Sousa$^{3}$ $^{\href{https://orcid.org/0000-0001-9047-2965}{\includegraphics[scale=0.5]{Images/orcid.jpg}}}$,
G. Olofsson$^{4}$ $^{\href{https://orcid.org/0000-0003-3747-7120}{\includegraphics[scale=0.5]{Images/orcid.jpg}}}$,\newauthor
M. Beck$^{5}$ $^{\href{https://orcid.org/0000-0003-3926-0275}{\includegraphics[scale=0.5]{Images/orcid.jpg}}}$,
N. Billot$^{5}$ $^{\href{https://orcid.org/0000-0003-3429-3836}{\includegraphics[scale=0.5]{Images/orcid.jpg}}}$,
S. Hoyer$^{6}$ $^{\href{https://orcid.org/0000-0003-3477-2466}{\includegraphics[scale=0.5]{Images/orcid.jpg}}}$,
S. Gill$^{7}$, 
N. Heidari$^{6}$,
D. V. Martin$^{8}$ $^{\href{https://orcid.org/0000-0002-7595-6360}{\includegraphics[scale=0.5]{Images/orcid.jpg}}}$,
C. M. Persson$^{9}$,
M. R. Standing$^{1}$ $^{\href{https://orcid.org/0000-0002-7608-8905}{\includegraphics[scale=0.5]{Images/orcid.jpg}}}$,\newauthor
Y. Alibert$^{10}$ $^{\href{https://orcid.org/0000-0002-4644-8818}{\includegraphics[scale=0.5]{Images/orcid.jpg}}}$,
R. Alonso$^{11,12}$ $^{\href{https://orcid.org/0000-0001-8462-8126}{\includegraphics[scale=0.5]{Images/orcid.jpg}}}$,
G. Anglada$^{13,14}$ $^{\href{https://orcid.org/0000-0002-3645-5977}{\includegraphics[scale=0.5]{Images/orcid.jpg}}}$,
J. Asquier$^{15}$,
T. Bárczy$^{16}$ $^{\href{https://orcid.org/0000-0002-7822-4413}{\includegraphics[scale=0.5]{Images/orcid.jpg}}}$,
D. Barrado$^{17}$ $^{\href{https://orcid.org/0000-0002-5971-9242}{\includegraphics[scale=0.5]{Images/orcid.jpg}}}$,\newauthor
S. C. C. Barros$^{3,18}$, $^{\href{https://orcid.org/0000-0003-2434-3625}{\includegraphics[scale=0.5]{Images/orcid.jpg}}}$,
M. P. Battley$^{7}$,
W. Baumjohann$^{19}$, $^{\href{https://orcid.org/0000-0001-6271-0110}{\includegraphics[scale=0.5]{Images/orcid.jpg}}}$,
T. Beck$^{10}$,
W. Benz$^{10,20}$, $^{\href{https://orcid.org/0000-0001-7896-6479}{\includegraphics[scale=0.5]{Images/orcid.jpg}}}$,
M. Bergomi$^{21}$,\newauthor
I. Boisse$^{6}$ $^{\href{https://orcid.org/0000-0002-3586-1316}{\includegraphics[scale=0.5]{Images/orcid.jpg}}}$,
X. Bonfils$^{22}$ $^{\href{https://orcid.org/0000-0001-9003-8894}{\includegraphics[scale=0.5]{Images/orcid.jpg}}}$,
A. Brandeker$^{4}$ $^{\href{https://orcid.org/0000-0002-7201-7536}{\includegraphics[scale=0.5]{Images/orcid.jpg}}}$,
C. Broeg$^{10,20}$ $^{\href{https://orcid.org/0000-0001-5132-2614}{\includegraphics[scale=0.5]{Images/orcid.jpg}}}$,
J. Cabrera$^{23}$,
S. Charnoz$^{24}$ $^{\href{https://orcid.org/0000-0002-7442-491X}{\includegraphics[scale=0.5]{Images/orcid.jpg}}}$,\newauthor
A. Collier Cameron$^{25}$ $^{\href{https://orcid.org/0000-0002-8863-7828}{\includegraphics[scale=0.5]{Images/orcid.jpg}}}$,
Sz. Csizmadia$^{23}$ $^{\href{https://orcid.org/0000-0001-6803-9698}{\includegraphics[scale=0.5]{Images/orcid.jpg}}}$,
M. B. Davies$^{26}$ $^{\href{https://orcid.org/0000-0001-6080-1190}{\includegraphics[scale=0.5]{Images/orcid.jpg}}}$,
M. Deleuil$^{6}$ $^{\href{https://orcid.org/0000-0001-6036-0225}{\includegraphics[scale=0.5]{Images/orcid.jpg}}}$,
L. Delrez$^{27,28}$ $^{\href{https://orcid.org/0000-0001-6108-4808}{\includegraphics[scale=0.5]{Images/orcid.jpg}}}$,\newauthor
O. D. S. Demangeon$^{3,18}$ $^{\href{https://orcid.org/0000-0001-7918-0355}{\includegraphics[scale=0.5]{Images/orcid.jpg}}}$,
B.-O. Demory$^{20}$ $^{\href{https://orcid.org/0000-0002-9355-5165}{\includegraphics[scale=0.5]{Images/orcid.jpg}}}$,
G. Dransfield$^{1}$ $^{\href{https://orcid.org/0000-0002-3937-630X}{\includegraphics[scale=0.5]{Images/orcid.jpg}}}$,
D. Ehrenreich$^{5}$ $^{\href{https://orcid.org/0000-0001-9704-5405}{\includegraphics[scale=0.5]{Images/orcid.jpg}}}$,
A. Erikson$^{23}$,\newauthor
A. Fortier$^{10,20}$ $^{\href{https://orcid.org/0000-0001-8450-3374}{\includegraphics[scale=0.5]{Images/orcid.jpg}}}$,
L. Fossati$^{19}$ $^{\href{https://orcid.org/0000-0003-4426-9530}{\includegraphics[scale=0.5]{Images/orcid.jpg}}}$,
M. Fridlund$^{29,9}$ $^{\href{https://orcid.org/0000-0002-0855-8426}{\includegraphics[scale=0.5]{Images/orcid.jpg}}}$,
D. Gandolfi$^{30}$ $^{\href{https://orcid.org/0000-0001-8627-9628}{\includegraphics[scale=0.5]{Images/orcid.jpg}}}$,
M. Gillon$^{27}$ $^{\href{https://orcid.org/0000-0003-1462-7739}{\includegraphics[scale=0.5]{Images/orcid.jpg}}}$,
M. Güdel$^{31}$,
J. Hasiba$^{19}$,\newauthor
G. Hébrard$^{32}$,
K. Heng$^{20,7}$ $^{\href{https://orcid.org/0000-0003-1907-5910}{\includegraphics[scale=0.5]{Images/orcid.jpg}}}$,
K. G. Isaak$^{33}$ $^{\href{https://orcid.org/0000-0001-8585-1717}{\includegraphics[scale=0.5]{Images/orcid.jpg}}}$,
L. L. Kiss$^{34,35}$,
E. Kopp$^{23}$,
V. Kunovac$^{48,1}$ $^{\href{https://orcid.org/0000-0001-9419-3736}{\includegraphics[scale=0.5]{Images/orcid.jpg}}}$,
J. Laskar$^{36}$ $^{\href{https://orcid.org/0000-0003-2634-789X}{\includegraphics[scale=0.5]{Images/orcid.jpg}}}$,\newauthor
A. Lecavelier des Etangs$^{32}$ $^{\href{https://orcid.org/0000-0002-5637-5253}{\includegraphics[scale=0.5]{Images/orcid.jpg}}}$,
M. Lendl$^{5}$ $^{\href{https://orcid.org/0000-0001-9699-1459}{\includegraphics[scale=0.5]{Images/orcid.jpg}}}$,
C. Lovis$^{5}$ $^{\href{https://orcid.org/0000-0001-7120-5837}{\includegraphics[scale=0.5]{Images/orcid.jpg}}}$,
D. Magrin$^{21}$ $^{\href{https://orcid.org/0000-0003-0312-313X}{\includegraphics[scale=0.5]{Images/orcid.jpg}}}$,
J. McCormac$^{7}$,
N. J. Miller$^{2}$ $^{\href{https://orcid.org/0000-0001-9550-1198}{\includegraphics[scale=0.5]{Images/orcid.jpg}}}$,\newauthor
V. Nascimbeni$^{21}$ $^{\href{https://orcid.org/0000-0001-9770-1214}{\includegraphics[scale=0.5]{Images/orcid.jpg}}}$,
R. Ottensamer$^{31}$ $^{\href{https://orcid.org/0000-0001-5684-5836}{\includegraphics[scale=0.5]{Images/orcid.jpg}}}$,
I. Pagano$^{38}$ $^{\href{https://orcid.org/0000-0001-9573-4928}{\includegraphics[scale=0.5]{Images/orcid.jpg}}}$,
E. Pallé$^{11}$ $^{\href{https://orcid.org/0000-0003-0987-1593}{\includegraphics[scale=0.5]{Images/orcid.jpg}}}$,
F. A. Pepe$^{5}$,
G. Peter$^{39}$ $^{\href{https://orcid.org/0000-0001-6101-2513}{\includegraphics[scale=0.5]{Images/orcid.jpg}}}$,
G. Piotto$^{21,40}$ $^{\href{https://orcid.org/0000-0002-9937-6387}{\includegraphics[scale=0.5]{Images/orcid.jpg}}}$,\newauthor
D. Pollacco$^{7}$,
D. Queloz$^{41,42}$, $^{\href{https://orcid.org/0000-0002-3012-0316}{\includegraphics[scale=0.5]{Images/orcid.jpg}}}$,
R. Ragazzoni$^{21,40}$ $^{\href{https://orcid.org/0000-0002-7697-5555}{\includegraphics[scale=0.5]{Images/orcid.jpg}}}$,
N. Rando$^{15}$,
H. Rauer$^{23,43,44}$ $^{\href{https://orcid.org/0000-0002-6510-1828}{\includegraphics[scale=0.5]{Images/orcid.jpg}}}$,
I. Ribas$^{13,14}$ $^{\href{https://orcid.org/0000-0002-6689-0312}{\includegraphics[scale=0.5]{Images/orcid.jpg}}}$,\newauthor
S. Lalitha$^{1}$ $^{\href{https://orcid.org/0000-0001-8102-3033}{\includegraphics[scale=0.5]{Images/orcid.jpg}}}$,
A. Santerne$^{6}$,
N. C. Santos$^{3,18}$ $^{\href{https://orcid.org/0000-0003-4422-2919}{\includegraphics[scale=0.5]{Images/orcid.jpg}}}$,
G. Scandariato$^{38}$ $^{\href{https://orcid.org/0000-0003-2029-0626}{\includegraphics[scale=0.5]{Images/orcid.jpg}}}$,
D. Ségransan$^{5}$ $^{\href{https://orcid.org/0000-0003-2355-8034}{\includegraphics[scale=0.5]{Images/orcid.jpg}}}$,
A. E. Simon$^{10}$ $^{\href{https://orcid.org/0000-0001-9773-2600}{\includegraphics[scale=0.5]{Images/orcid.jpg}}}$,\newauthor
A. M. S. Smith$^{23}$ $^{\href{https://orcid.org/0000-0002-2386-4341}{\includegraphics[scale=0.5]{Images/orcid.jpg}}}$,
M. Steller$^{19}$ $^{\href{https://orcid.org/0000-0003-2459-6155}{\includegraphics[scale=0.5]{Images/orcid.jpg}}}$,
Gy. M. Szabó$^{45,46}$,
N. Thomas$^{10}$,
S. Udry$^{5}$ $^{\href{https://orcid.org/0000-0001-7576-6236}{\includegraphics[scale=0.5]{Images/orcid.jpg}}}$,
V. Van Grootel$^{28}$ $^{\href{https://orcid.org/0000-0003-2144-4316}{\includegraphics[scale=0.5]{Images/orcid.jpg}}}$,\newauthor
N. A. Walton$^{47}$ $^{\href{https://orcid.org/0000-0003-3983-8778}{\includegraphics[scale=0.5]{Images/orcid.jpg}}}$\\
Affiliations are listed at the end of the paper.
}

\date{Accepted. Received; in original form}

\pubyear{2022}

\begin{document}
\label{firstpage}
\pagerange{\pageref{firstpage}--\pageref{lastpage}}
\maketitle

\begin{abstract} %(max 250 words)
Eclipsing binaries are important benchmark objects to test and calibrate stellar structure and evolution models. This is especially true for binaries with a fully convective M-dwarf component for which direct measurements of these stars' masses and radii are difficult using other techniques. Within the potential of M-dwarfs to be exoplanet host stars, the accuracy of theoretical predictions of their radius and effective temperature as a function of their mass is an active topic of discussion. Not only the parameters of transiting exoplanets but also the success of future atmospheric characterisation rely on accurate theoretical predictions.
We present the analysis of five eclipsing binaries with low-mass stellar companions out of a sub-sample of 23, for which we obtained ultra high-precision light curves using the \textit{CHEOPS} satellite. The observation of their primary and secondary eclipses are combined with spectroscopic measurements to precisely model the primary parameters and derive the M-dwarfs mass, radius, surface gravity, and effective temperature estimates using the \texttt{PYCHEOPS} data analysis software. 
Combining these results to the same set of parameters derived from \textit{TESS} light curves, we find very good agreement (better than 1\% for radius and better than 0.2\% for surface gravity).
We also analyse the importance of precise orbits from radial velocity measurements and find them to be crucial to derive M-dwarf radii in a regime below 5\% accuracy. 
These results add five valuable data points to the mass-radius diagram of fully-convective M-dwarfs.

\end{abstract}

\begin{keywords}
binaries: eclipsing -- stars: fundamental parameters -- stars: low-mass -- techniques: photometric -- techniques: spectroscopic
\end{keywords}

%%%%%%%%%%%%%%%%%%%%%%%%%%%%%%%%%%%%%%%%%%%%%%%%%%

%%%%%%%%%%%%%%%%% BODY OF PAPER %%%%%%%%%%%%%%%%%%

\section{Introduction}

Low-mass main-sequence stars of M-type (M-dwarfs) have been in the spotlight of recent exoplanet surveys \citep{Nutzman08,Delrez18,Barclay18,quirrenbach19,Donati20}. 
This development has two main reasons. First their low masses, and radii, compared to F, G, and K stars make it easier to detect small planets and planetary systems composed of mini Neptunes down to Earth sized planets by means of radial velocity and transit methods \citep[e.g.][]{gillon16, zechmeister19, guenther19}. Thus, more Earth sized planets have been found in the habitable zone of M-dwarfs than for solar-type stars \citep[e.g.][]{Dressing13}. Second, M-dwarfs have low luminosities and, thus offer the first possible window to study transiting rocky planets in their habitable zone and directly analyse their atmospheres with high-precision instruments like the James Webb Space telescope \citep{Kaltenegger09,Morley17}.

Such studies depend crucially on the knowledge of the parameters of M-dwarf planets which in turn are derived from the mass and radius of the host M-dwarf. 
Up to now our understanding on the mass and radius distribution of low-mass stars which are fully convective ($\rm M_{\star} < 0.35\,M_{\sun}$, \citealt{Chabrier97}) is rather poorly explored compared to more massive stars. This is mainly due to the relative faintness of these stars\footnote{E.g. the planet host star TRAPPIST-1, a M7.5 ultra-cool dwarf in 12\,pc distance has a visual magnitude of only 18.8\,mag .}. Especially the lack of a large sample of M-dwarfs with directly measured mass and radius make it difficult to calibrate stellar evolution models which are typically used to estimate the properties of planet host stars like for example the Exeter/Lyon models \citep{Baraffe15} or the Dartmouth models \citep{Dotter08}.

Studies of M-stars with available radii and masses have revealed that their stellar radii for a given mass are apparently inflated by a few percent, compared to estimates from models \citep[e.g.][]{Casagrande08,Torres10,Spada13,Kesseli18}. 

Several possible explanations have been discussed, like stellar magnetic activity \citep{mullan01,Chabrier07}, or a bias due to binarity \citep{ribas06,Morales09}. Also metallicity effects seem to play a role \citep{Berger06,boetticher19}.
%The metallicity-dependend sample from \cite{boetticher19} suggests, that an uncertainty on 0.2\,dex could lead to a radius uncertainty of \sim 2.4\%. 
Thus, a representative sample of low-mass M-dwarfs with accurately measured mass, radius, but also metallicity is crucial to understand how the different effects enter into this radius inflation problem. 

%EBLM - short into

The eclipsing binaries with low mass (EBLM) project \citep{triaud13} is focusing on a large sample with hundreds of eclipsing binaries of F,G, \& K-type stars, orbited by late type M-dwarf companions. These binaries have been detected from the WASP survey \citep{pollacco06}. Using a large radial velocity follow-up campaign of these stars, \cite{triaud17} derived accurate orbits of many of these systems thus being able to measure fundamental parameters like precise mass and radius of the low-mass M-dwarfs. The binary configuration with a solar-type star allows us to measure accurately the metallicity of the solar-type star. Assuming an equal metallicity of both components, we can constrain the metallicity of the M-dwarf.  Thus, EBLM targets are ideal candidates to populate the mass regime of fully convective M-dwarfs with masses below $\rm 0.35\,M_{\sun}$ and to establish an empirical mass-radius-metallicity relationship for these stars. Early results from sub samples indicate that models can be matched quite well, when taking accurate measurements of the metallicty of the M-dwarf into account \citep{boetticher19,Gill19}. Every low-mass M-dwarf with accurately measured mass, radius and metallicity will help to tighten the constraints on the source of the radius inflation problem and in return will allow us in future to constrain precise parameters of planet host stars.         

%CHEOPS - short intro

\textit{CHEOPS} \citep{benz21} is a S-class mission of the European Space Agency, which has been launched on the 18th of December 2019. Its primary mission is to perform ultra high-precision photometry of bright exoplanet host-stars. We have started an `Ancillary Science' programme on a selection of 23 EBLM targets, to obtain precise measurements of primary and secondary eclipses, which allow us to (i) derive the size of both components and (ii) to measure the M-dwarf effective temperature from the surface brightness ratios. Additionally, we use light curves, obtained by the \textit{TESS} survey \citep{ricker15}, which covers the northern and southern hemispheres with observing periods of about one month per pointing (sector). \textit{TESS} cameras have a three times smaller aperture compared to \textit{CHEOPS}, leading to a lower accuracy for eclipse events in \textit{TESS} data. Nevertheless, the long coverage of photometric data allows us to gather multiple eclipses of our targets and thus improve and compare orbital parameters, as well as to optimise our analysis of \textit{CHEOPS} observations.

The three EBLM binaries, analysed in our \textit{CHEOPS} programme EBLM\,J1741+31, EBLM\,J1934-42 and EBLM\,J2046+06 have shown that M-dwarfs with precisely measured radii and metallicities open up the possibility to disentangle the effect of metallicity from different effects on the radius inflation problem for low-mass M-dwarfs \citep{swayne21}.

In this paper we present the analysis of five EBLM binaries with fully convective M-dwarfs companions, observed in our \textit{CHEOPS} programme and compare them to the analysis of \textit{TESS} observations.

\begin{table*}
	\centering
	\caption{\textit{CHEOPS} observations and data extraction for our targets. Effic. is the fraction of the observation that resulted in valid (usable) data and $\rm R_{ap}$ the aperture radius used to extract the light curves.}
	\label{tab:CHEOPS_log}
	\begin{tabular}{lccccccc}
    \hline
    Eclipse & Target  & Start date & Duration & $\rm T_{exp} $ & Effic. & File key & $\rm R_{ap}$ \\
    Event   &         & (UTC) & (h) & (s) & (\%) & & (pixels) \\ 
    \hline
    Primary & EBLM J0239-20  & 2020-11-01T15:43 & 8.80 & 60 & 86.2 & CH\_PR100037\_TG012001\_V0200 & 25  \\
    Secondary &         & 2020-11-05T20:30 & 7.99 & 60 & 93.2 & CH\_PR100037\_TG011901\_V0200 & 25 \\
    Secondary &         & 2020-11-19T17:24 & 9.02 & 60 & 70.4 & CH\_PR100037\_TG011902\_V0200 & 25 \\
    \hline
    Primary & EBLM J0540-17  & 2020-12-07T08:39 & 10.04 & 60 & 68.4 & CH\_PR100037\_TG012601\_V0200 & 17.5 \\
    Secondary &         & 2021-01-21T09:39 & 10.75 & 60 & 54.1 & CH\_PR100037\_TG012502\_V0200 & 17.5 \\
    Secondary &         & 2020-12-04T08:13 & 10.62 & 60 & 66.5 & CH\_PR100037\_TG012501\_V0200 & 17.5 \\
    Secondary &         & 2021-01-27T09:20 & 10.49 & 60 & 50.0 & CH\_PR100037\_TG012503\_V0200 & 17.5 \\
    \hline
    Primary & EBLM J0546-18    & 2020-11-30T22:27 & 8.67 & 60 & 67.5 & CH\_PR100037\_TG012801\_V0200 & 25\\
    Secondary &           & 2020-12-31T05:35 & 8.77 & 60 & 66.3 & CH\_PR100037\_TG012701\_V0200 & 25\\
    Secondary &           & 2021-01-09T19:50 & 8.05 & 60 & 64.0 & CH\_PR100037\_TG012702\_V0200 & 25\\
    \hline
    Primary & EBLM J0719+25    & 2020-12-10T07:03 & 8.80 & 60 & 52.8 & CH\_PR100037\_TG013001\_V0200 & 22.5\\
    Secondary &           & 2021-02-03T20:54 & 8.69 & 60 & 57.7 & CH\_PR100037\_TG017301\_V0200 & 22.5\\
    Secondary$^1$ &     & 2020-12-21T12:03 & 8.50 & 60 & 60.2 & CH\_PR100037\_TG012901\_V0200 & 22.5\\
    \hline
    Secondary & EBLM J2359+44    & 2020-11-11T08:59 & 8.89 & 60 & 58.3 & CH\_PR100037\_TG016301\_V0200 & 26.5 \\
    Primary &           & 2020-11-28T13:07 & 15.67& 60 & 51.4 & CH\_PR100037\_TG016401\_V0200 & 26.5 \\
    \hline
    \end{tabular}
    \\
    $^1$ For this observation the secondary eclipse of EBLM J0719+25 has been missed, thus we cannot use this data set for parameter determination of the binary.
\end{table*}

\section{Observations and Methods}

Primary and secondary eclipses for all our five eclipsing binaries were observed with \textit{CHEOPS} between November 2020 and January 2021 as part of \textit{CHEOPS} Guaranteed Time Observation programme ID-037. We obtained one primary eclipse and, depending on the depth of the secondary eclipse, one to three secondary eclipse observations in order to obtain sufficient signal to noise to measure both eclipses. Table~\ref{tab:CHEOPS_log} gives an overview of the \textit{CHEOPS} observations and data extraction. All data were reduced by the \textit{CHEOPS} data reduction pipeline v13.1 \citep{hoyer20}, which performs an aperture photometry of the target star, taking contamination in the field as well as instrumental effects like the rotation of the satellite into account. The pipeline offers light curves for different aperture sizes. For our analysis, we selected the aperture size with minimal median absolute deviation of the point-to-point difference in the light curve. The resulting aperture radii are listed as $\rm R_{ap}$ in Table~\ref{tab:CHEOPS_log}. The observations were interrupted due to the low-Earth-orbit of \textit{CHEOPS} by Earth occultations, as well as crossings of the South Atlantic Anomaly. We derive the time spent on target as the fraction of valid observations compared to the total observation interval.

The \textit{TESS} survey covered all of our targets with 2-min cadence data made available by \textit{TESS} Guest Investigator (GI) programmes. 
EBLM J0239-20 (TIC64108432) has been observed in sectors 4 and 31 under GI programmes G011278 and G03216. EBLM J0540-17 (TIC46627823) has been observed in sectors 6 and 32 under GI programmes G011278, G03216, \& G03251. EBLM J0546-18 (TIC93334206) has been observed in sectors 32 and 33 under GI programme G03216. EBLM J0719+25 (TIC458377744) has recently been observed in sectors 44, 45, \& 46 under GI programme G04157 and EBLM J2359+44 (TIC177644756) has been observed in sector 17 under GI programmes G022253 \& G022156.  
Data reduction and light curve extraction were done by the \textit{TESS} Science Processing Operations Center Pipeline (SPOC; \citealt{jenkins16}) and were downloaded via the Mikulski Archive for Space Telescopes\footnote{\url{https://archive.stsci.edu/}} (MAST). For our analysis, we used Pre-search Data Conditioned Simple Aperture Photometry (PDCSAP) flux data and bitmask 175 to exclude data flagged with severe quality issues \citep{tenenbaum18}.

For EBLM J2359+44 two radial velocity measurements have been published by \cite{Poleski10} that confirmed it to be a binary star. Full time series radial velocity observations of EBLM J0719+25 and EBLM J2359+44 have been taken with the SOPHIE high-resolution echelle spectrograph \citep{perruchot08}, mounted on the 1.93\,m telescope at the Observatoire de Haute-Provence in France as part of the Binaries Escorted By Orbiting Planets (BEBOP) survey to search for circumbinary planets \citep{martin19}.
For EBLM J0719+25, 8 SOPHIE spectra have been obtained between November 2018 and October 2019 in High-Resolution mode (R = 75 000). For EBLM J2359+44, 15 SOPHIE spectra have been obtained between November 2018 and September 2020 in High-Resolution mode (R = 75 000) as well as in High-Efficiency  (HE) mode (R = 40 000). The HE mode allows an about 2.5 times higher throughput compared to the High-Resolution mode. The spectra have an average signal to noise of about 30 with a typical exposure time of 1800\,s. To allow the removal of the background contamination from the Moon, all observations were taken with one fibre on target and one on the sky. The spectra were reduced using the SOPHIE Data Reduction Software \citep{baranne96} and radial velocities were measured by cross correlation with a G2 mask \citep{Courcol15} for which we achieved a typical precision of ~10\,m\,$\rm s^{-1}$ for our spectra. All radial velocity measurements are listed in the Appendix Tables\, \ref{tab:rv_J0719+25} \&\,\ref{tab:rv_J2359+44}.
We submitted a target list of 40 EBLM systems from \cite{triaud17} as a priority 4 proposal to be observed with high resolution spectrograph \citep{2014SPIE.9147E..6TC} of the Southern African Large Telescope (SALT) in medium resolution ($R \approx 37,000$). In total, 30 of them were observed between the 19th of May and 7th August 2017, including EBLM J0239-20. These observations were made in long slit mode with an exposure time scaling as a function of magnitude to ensure a SNR $\geq \, 100$. Data was reduced and processed using standard pipelines \citep{2015ascl.soft11005C, 2015ascl.soft10007C} to produce two spectra for each observation (370--550 \,nm \& 550--890\,nm) as a result of the dual-beam nature of the spectrograph.
%The radial velocities, obtained from SOPHIE observation of a few meters per seconds (e.g. Bouchy et al. 2013; Hara et al. 2020)

\begin{table*}
	\centering
	\caption{Stellar and orbital parameters of the primary stars. Coordinates are in J2000.}
	\label{tab:star_params}
	\begin{tabular}{lccccc}
    \hline
    					& EBLM J0239-20 	    & EBLM J0540-17 	& EBLM J0546-18     & EBLM J0719+25 		& EBLM J2359+44 \\
    					\hline
    Name   				& TYC 5862-1683-1  	    & TYC 5921-745-1 	&  TIC 93334206 	& TYC1913-0843-1 		& TYC3245-0077-1\\
    RA      		    & 02 39 29.29  			& 05 40 43.58 		& 05 46 04.81  		& 07 19 14.26 			& 23 59 29.74 \\
    Dec.    		    & $-$20 02 24.0  			& $-$17 32 44.8 		& $-$18 17 54.6    	& +25 25 30.8 			& +44 40 31.2 \\
    G (mag)		        & 10.57 				& 11.42 			& 12.01		 	    & 11.15 				& 10.46 \\
    Sp. Type			    & G0 				& G0 				& G0 			    & G0 			    	& F8 \\
    $\rm T_{eff,1} (K)^a$ 	& 5758 $\pm$ 100 	& 6290 $\pm$ 77 	& 6180 $\pm$ 80		& 6026 $\pm$ 67			& 6799 $\pm$ 83\\
    $\rm log\,g_1 (cgs)^c$ 	& 4.053 $\pm$ 0.016 & 4.058 $\pm$ 0.017	& 4.100 $\pm$ 0.034	& 4.239 $\pm$ 0.022		& 4.068 $\pm$ 0.010\\
    $\rm [Fe/H]^a$ 		& 0.27 $\pm$ 0.12 		& $-$0.04 $\pm$ 0.05 	& $-$0.45 $\pm$ 0.08 	& 0.04 $\pm$ 0.05		& 0.12 $\pm$ 0.05\\
    $\rm R_1 (R_{\odot})^c$ &1.587 $\pm$ 0.039 	& 1.636 $\pm$ 0.040	& 1.509 $\pm$ 0.064	& 1.305 $\pm$ 0.038		& 1.711 $\pm$      0.033\\
    $\rm M_1 (M_{\odot})^c$ &1.037 $\pm$ 0.060	& 1.120 $\pm$ 0.062	& 1.051 $\pm$ 0.059	& 1.078 $\pm$ 0.059 	& 1.253 $\pm$      0.070\\
    \\%$\rm \rho_1$			& 
    Orbital parameters:\\
    $\rm K (km\,s^{-1})$ 	& 21.316$\pm$0.036$\rm^d$	& 16.199$\pm$0.010$\rm^d$	& 26.15$\pm$0.10$\rm^d$	& 15.02$\pm$0.04$\rm^b$	& 23.62$\pm$0.08$\rm^b$\\
    $\rm e$                 & $\rm <0.0032^d$                & $\rm 0.00029\pm0.00057^d$                & $\rm <0.015^d$            & 0.0730$\pm$0.0045$\rm^b$& 0.4773$\pm$0.0010$\rm^b$\\
    $\rm \omega (deg)$      & --                        & $\rm -164\pm10^d$                        & --                    & $-$155.8$\pm$5.4$\rm^b$  & $-$94.290$\pm$0.060$\rm^b$\\
    $f\!(m)\,(10^{-3} M_{\odot})$                  & 2.788$\pm$0.014$\rm^d$ & 2.6444$\pm$0.0096$\rm^d$ & 2.1332$\pm$0.0023$\rm^d$ & 2.597$\pm$0.021$\rm^b$ & 10.53$\pm$0.11$\rm^b$ \\
    \hline
    \end{tabular}
    \\
    References:
    $\rm^a$ From spectral analysis,$\rm^b$ from radial velocity analysis, $\rm^c$ from light curve modelling, $\rm^d$ from \cite{triaud17}\\
\end{table*}

\begin{table}
	\centering
	\caption{Priors on $f_c = \sqrt{e}\cos{\omega}$ and $f_s = \sqrt{e}\sin{\omega}$ used in the analysis of the \textit{CHEOPS} and \textit{TESS} light curves based on the spectroscopic orbits for each binary system.}
	\label{tab:fit_priors}
	\begin{tabular}{lccccc}
    \hline
    Target      &       $\rm f_c$ & $\rm f_s$ \\\hline
    EBLM J0239-20    & 0.0   & 0.0 \\
    EBLM J0540-17    & 0.0   & 0.0 \\
    EBLM J0546-18    & 0.0   & 0.0 \\
    EBLM J0719+25    & $-$0.247$\pm$0.013 & $-$0.111$\pm$0.023 \\
    EBLM J2359+44    & $-$0.0517$\pm$0.0007 & $-$0.6889$\pm$0.0007 \\
    
    %CHEOPS fit \\
    
    %$\rm h_1$ & & 0.766 $\pm$ 0.012 &&& \\
    %$\rm h_2$ & & 0.450 $\pm$ 0.050 &&& \\
    \hline
    \end{tabular}
    \\
    
\end{table}

\section{Analysis}
For data analysis, we followed the methods, described in \cite{swayne21}, hereafter SW21. Both \textit{TESS} and \textit{CHEOPS} light curves were modelled using the \texttt{qpower2} transit model, which applies a power-2 limb darkening law \citep{maxted19}. We use it as binary star model including primary and secondary eclipses which is implemented in the python software \texttt{PYCHEOPS}\footnote{\url{https://github.com/pmaxted/pycheops}} \citep{maxted21}. The parameters of the binary star model are the orbital period $\rm P$, the mid-time of the primary eclipse $\rm T_{0}$; the primary and secondary eclipse depths $\rm D$ and $\rm L$, the impact parameter $\rm b$, the parameters $\rm f_c = \sqrt{e}\,cos(\omega)$ and $\rm f_s = \sqrt{e}\,sin(\omega)$, which parameterise the eccentricity $\rm e$ and the longitude of periastron $\rm \omega$, the limb darkening parameters $\rm h_1$ and $\rm h_2$ \citep{maxted18}, and $\rm W$, which becomes the width of the eclipse for e = 0 and is defined by the stellar radii, impact parameter, and the semi mayor axis $\rm a$ (see \cite{maxted21} for details).
%$\rm W = \sqrt{(1+k)^2 - b^2}\,R_1/(\pi\,a)$, W becomes the   %= R_^21 / R_1^2$,
We used gaussian priors for $\rm f_c, f_s$. These priors were derived from radial velocity measurements of the systems. Orbital parameters from radial velocity measurements for EBLM J0239-20, EBLM J0540-17, and EBLM J0546-18 have been published in \cite{triaud17}. Their eccentricities are reported to be consistent to zero, thus we set those priors to zero for all three systems. For EBLM J0719+25 and EBLM J2359+44, we used the binary star python code \texttt{ellc} \citep{maxted16}, to model the radial velocity from SOPHIE measurements as well as the two measurements from \cite{Poleski10} for EBLM J2359+44. We sampled the posterior probability distribution (PPD) of our model parameters $\rm f_c, f_s$, and the semi amplitude K, using the Markov chain Monte Carlo (MCMC) code \texttt{EMCEE} \citep{Foreman-Mackey13} to take the RV-jitter of the data into account by weighting the fit by the log-likelihood function. For this we used the period from our TESS fit (see Sec.\,\ref{tess_fit}) as fixed prior and did not need to fit any additional trend to the data. %The resulting jitter was 74\,$\rm m\,s^{-1}$ and 10\,$\rm m\,s^{-1}$ for EBLM J0719+25 and EBLM J2359+44 respectively. 
The resulting orbital parameters, as well as the mass function $f\!(m)$ (see equation 6 in \citealt{triaud17}) are listed in Table\,\ref{tab:star_params}. The resulting priors for $\rm f_c, f_s$ are listed in Table\,\ref{tab:fit_priors}. The errors represent the one sigma error of the resulting PPD.

%From these we derived the priors $\rm f_c = -0.247 \pm 0.013, f_s = -0.111 \pm 0.023$, and $\rm f_c = -0.0517 \pm 0.0007, f_s = -0.6889 \pm 0.0007$ for EBLM J0719+25 and EBLM J2359+44 respectively.

%, as well as the semi mayor axis to derive the light travel time correction for the secondary eclipse.
%%, and $\rm a = 13.58\,R_{1}$., $\rm a = 14.26\,R_{1}$.
%\ds{$\rm Used priors for CHEOPS:
%0239-20: fc,fs =0, a_c - EBLM IV
%0540-17: h1 h2 stagger using TESS stellar params, fc,fs - EBLM IV
%0546-18: fc,fs=0, aR -> light travel time correction??? - EBLM IV
%0719+25: fc, Fs, own fit, aR, L ->
%use always gaussian parameters between 0 and 1
%Used priors for TESS:
%0239-20: fc,fs =0, a\_c - EBLM IV , h1,h2 use always stagger as guessing parameters
%$}

\subsection{\textit{TESS} light curve analysis} \label{tess_fit}
Only segments of the \textit{TESS} light curve within one eclipse duration of the time of mid-eclipse were used in this analysis. To remove trends in the light curve, we divided these segments by a linear polynomial model fitted to the data either side of the eclipse. Unlike SW21, we preferred this method over the use of a Gaussian process in order to securely preserve the transit shape of the faint secondary eclipses. 

%we divide the light curve by a model of the variations. For this, we first bin the TESS light curve to 1\,h bins to only account for slowly varying flux variations. We then mask the eclipse events and model the light curve using the \texttt{CELERITE} software to apply a Gaussian process with a stochastically driven damped simple harmonic oscillator function \cite{Foreman-Mackey17}. In the case of EBLM J2359+44 we encountered some instability of the Gaussian process and modelled the out of eclipse flux variations using a polynomial model of order 25.
To model the light curve, we first determined the initial orbital parameters using a least-squares fit and then sampled the PPD of our transit model using \texttt{EMCEE}. We placed normal priors on the orbital parameters $\rm f_c, f_s$, as listed in Table\,\ref{tab:fit_priors} as well as on the white noise, using the residual rms of the least-squares fit. %We placed Gaussian priors on the limb darkening parameters $\rm h_1$ and $\rm h_2$. These were derived by interpolating the tables for the TESS bandpass published in \cite{maxted18} using the stellar parameters $\rm T_{eff,1}, log\,g_1, and [Fe/H]$ as listed in Table \ref{tab:star_params}, and applying an offset ($\rm h_1 + 0.01$ and $\rm h_2 - 0.045$; \citep{maxted18}). The priors used for each target are listed in Table\,\ref{tab:fit_priors}. 
The resulting parameters from the \textit{TESS} light curves are detailed in Tables\,\ref{tab:target_params1},\,\ref{tab:target_params2}, \& \,\ref{tab:target_params3}. These represent the median of the PPD as well as the standard errors from the 15.9\% and 84.1\% percentile-points of the PPD. We show the resulting fits of all targets in the Appendix, Fig.\,\ref{fig:Tess_lcs1} and Fig.\,\ref{fig:Tess_lcs2}.

\subsection{\textit{CHEOPS} light curve analysis} \label{cheops_fit}
\textit{CHEOPS} light curves were analysed in two steps. First we analysed every visit separately to derive initial model parameters (see Table \ref{tab:CHEOPS_log} for an overview of all visits).
As described in detail in SW21, instrumental effects like roll angle, contamination, and background level can be represented using linear correlation parameters or for roll angle $\rm \phi, sin(\phi), cos(\phi), sin (2\phi)$, etc., which were iteratively selected\footnote{See Table~\ref{tab:decorr} for the decorrelation parameters selected for each visit}. The PPD of all model and decorellation parameters were sampled simultaneously using \texttt{EMCEE}. We used the same Gaussian priors for $\rm f_c$, and $\rm f_s$ as for the \textit{TESS} data and since we obtained single eclipse events, we fixed our transit model to accurately measured orbital period P, from the \textit{TESS} light curve fit. For secondary eclipses, we used priors on the parameters D, W and b, as derived from the primary eclipse of each target. 

%(For the individual data sets, decorrelation against roll angle is done by including coefficients dfdsinphi, dfdcosphi, dfdsin2phi etc. in the mode)

 %For EBLM J0546-18 we placed Gaussian priors $\rm h_1$ and $\rm h_2$ using the tabulated values for the Kepler passband from \cite{maxted18} similarly interpolated and with the same offsets, that we used for TESS light curves. For all other targets, 
%The eclipses can be used to constrain the limb darkening parameters well enough, thus we placed broad Gaussian priors of $\rm h_1 = 0.72[0.0,1.0]$ and $\rm h_2=0.67[0.0,1.0]$. We used the semi mayor axis from the known radial velocity orbits to derive the light travel time correction for the secondary eclipse. For EBLM J0719+25 and EBLM J2359+44 these are $\rm a = 13.58\,R_{1}$ and $\rm a = 14.26\,R_{1}$ respectively. 
%We used \texttt{PYCHEOPS} to first estimate the best parameters using a least-squares fit. 

In a second step, we were using a single MCMC to perform a 'multivisit' analysis including all visits for a specific target. We used the same priors as for the individual analysis as well as the results as input parameters and used the function \texttt{multivisit} of \texttt{PYCHEOPS} to sample the joint PPD with \texttt{EMCEE}. Hereby we used the implicit decorrelation method for instrumental trends as described in \cite{maxted21}, keeping the number of harmonic terms to its default ($\rm N_{roll}=3$).
The resulting parameters from the \textit{CHEOPS} light curves are detailed in Tables\,\ref{tab:target_params1},\,\ref{tab:target_params2}, \& \,\ref{tab:target_params3}. These represent the median of the PPD as well as the standard errors from the 15.9\% and 84.1\% percentile-points of the PPD. We show the resulting fits of all targets in the Appendix, Fig.\,\ref{fig:CHEOPS_lcs1}, Fig.\,\ref{fig:CHEOPS_lcs2}, and Fig.\,\ref{fig:CHEOPS_lcs3} and in Table~\ref{tab:decorr} the resulting decorrelation parameters from the multivisit analysis.

\subsection{Stellar parameters} \label{Sec:stel_param}

We used co-added high-resolution spectra to derive the stellar parameters of the primary components ($\rm T_{eff}$ and [Fe/H]). 
For EBLM J0540-17, we used co-added CORALIE spectra, obtained by \cite{triaud17} and available from the ESO science archive facility\footnote{\url{http://archive.eso.org/}} and co-added SOPHIE spectra for EBLM J0719+25 and EBLM J2359+44. The stellar parameters for these three targets were derived using the equivalent width method following the same methodology, model atmospheres, and line list as described in \cite{sousa14} and \cite{santos13}. In here we applied the \texttt{ARES} code \citep{sousa15}, as well as the \texttt{MOOG} radiative transfer code \citep{sneden12}, assuming ionisation and excitation equilibrium of iron lines. 
For EBLM J0546-18 we used co-added CORALIE spectra and applied a wavelet decomposition method where we compare the coefficients from a wavelet decomposition to those from a grid of model spectra. Those model spectra were synthesised using the code \texttt{SPECTRUM} \citep{1994AJ....107..742G}, MARCS model atmospheres \citep{2008A&A...486..951G} as well as the atomic line list version 5 of the Gaia ESO survey \citep{2015PhyS...90e4010H}. The method is detailed in \cite{gill18} and has been found to deliver robust measurements for effective temperature and metallicity for spectra with relatively low SNR ($\rm SNR \gtrapprox 40$). 
For EBLM J0239-20 we used the SALT spectra and modeled the stellar fundamental parameters using the software \texttt{SME}\footnote{\url{http://www.stsci.edu/~valenti/sme.html}} 
 \citep[Spectroscopy Made Easy;][]{vp96, pv2017} that computes synthetic spectra with atomic and molecular line data from
\texttt{VALD}\footnote{\url{http://vald.astro.uu.se}}  \citep{Ryabchikova2015} which is compared to the observations. We chose the stellar atmosphere grid Atlas12 \citep{Kurucz2013} and
modelled  $\rm T_{eff}$, $\rm log\,g_{1}$, abundances and $\rm v\,sin\,i$ one parameter at a time. Due to the high rotational velocity ($\rm v\,sin\,i = 31 \pm 4$~km~s$^{-1}$), the uncertainties in $\rm log\,g_{1}$ derived from the line wings of the \ion{Ca}{I} triplet around $\rm 6200~\AA$ is with $\rm 0.2\,dex$ relatively high. We thus rely on the lightcurve modelling to derive the surface gravity of our targets.

%We find  $T_\mathrm{eff} = 5758 \pm 100$~K, logg\,=\,$4.05 \pm 0.20$, [Fe/H]\,=\,$0.27\pm0.10$, [Ca/H]\,=\,$0.27\pm0.20$, and 
%and applied the \texttt{SME} code \citep{1996A&AS..118..595V,2017A&A...597A..16P} for spectral synthesis using the same model atmospheres and line list as  we used for the equivalent width method. 

% microturbulence~\rho_t, 
Similarly to SW21, we derived the system parameters using the function \texttt{massradius} in \texttt{PYCHEOPS}. 
As explained in \cite{maxted21}, this function applies a Monte Carlo approach to derive basic system parameters like the primaries mean stellar density, the mass and radius of the M-dwarf, using the PPD of our \textit{CHEOPS} light curve fit. 
It additionally uses the primaries mass and radius, as well as the orbital parameters which were not sampled in the PPD like period, and eccentricity as input and derives the surface gravity $\rm log\,g_{2}$ of the M-dwarf using the radial-velocity semi-amplitudes.% by applying formula (4) of \cite{}. 
We used this function to optimise the global system parameters in a two stage iterative process.

In the first step, we used the primaries mass and radius estimates available from the \textit{TESS} input catalogue v8 \citep{stassun19} as initial parameters. The derivation of these estimates is based on an empirical relation including photometric effective temperature estimates for stars with well measured Gaia distances. We used the same priors for period and eccentricity that we used for our \textit{CHEOPS} fit, as well as the semi-amplitudes from radial velocity measurements. For EBLM J0239-20, EBLM J0540-17, and EBLM J0546-18 we have used the published semi-amplitudes \citep{triaud17}, For For EBLM J0719+25 and EBLM J2359+44, we use the results from our orbital fit (see Table\,\ref{tab:star_params}).

In a second iteration, we made use of the \texttt{massradius} function again in order to find the best fitting parameters of the primary mass and radius from our light curve fit.
We used the relation of \cite{enoch10} (equation 4), to derive a mass sample for the primary star. This sample is based on the stellar density samples obtained from the first iteration and created similar sized samples for $\rm T_{eff}$ and [Fe/H] based on our spectroscopic stellar parameters. We then added  a normal distributed scatter of 0.023 to account for the resulting scatter for this relation found by \cite{enoch10}. We derived a radius sample using this mass sample as well as the density sample. We used the mass and radius samples to re-run the \texttt{massradius} function to derive the final stellar parameters of the primary and M-dwarf components. 
We finally derived the surface gravity $\rm log\,g_{1}$ from the stellar density, directly measured from the light curve fit of our \textit{CHEOPS} data, as well as the primaries mass derived from the previous step.

We derived the effective temperature T$\rm _{eff,2}$ of the M-dwarf companion using the surface brightness ratio $\rm L/D$, derived from the light curve fit of primary and secondary eclipses. 
Similar to  SW21, we derived the integrated surface brightness in the \textit{CHEOPS} and \textit{TESS} bandbands of the primary star, using the spectral parameters T$\rm _{eff,2}$, $\rm log\,g_{1}$, and [Fe/H] using PHOENIX model atmospheres with no alpha-element enhancement \citep{husser13} and sampled a large set of surface temperatures over the known parameters, L/D, log\,g, and [Fe/H] (assuming similar metallicity for both companions) to derive the effective temperature.

The light contribution from the primary star reflected to the M-dwarf can be expressed by $\rm A_{g}(R_{2}/a)^2$, where $\rm A_{g}$ is the geometric albedo and $\rm R_{2}/a$ is the radius of the M-dwarfs in units of the semi mayor axis, which we directly measure from our model. With a typical albedo of $\rm A_{g} \sim 0.1$ \citep{Marley99}, the light contribution for our targets is very small and thus negligible. Nevertheless, for the two shortest period binaries in our sample, EBLM J0239-20 and EBLM J0546-18 the light contribution might cause an underestimation of the secondary eclipse depth on the one sigma level and thus an underestimation of T$\rm _{eff,2}$ in the order of 1\,\% for both CHEOPS and TESS passbands. Thus, we increased the relative uncertainties for T$\rm _{eff,2}$ for EBLM J0239-20 and EBLM J0546-18 by 1\% in order to account for the unknown uncertainty of $\rm A_{g}$.

All parameters of the primary stars are listed in Table \ref{tab:star_params}, all parameters for the M-dwarf companions are listed in Tables\,\ref{tab:target_params1},\,\ref{tab:target_params2}, \&\,\ref{tab:target_params3}.

\begin{table*}
	\centering
	\caption{The derived parameters for EBLM J0239-20 and EBLM J0540-17 using \textit{CHEOPS} and \textit{TESS} light curve fits with eclipse depths being in the relevant instrumental bandpass.}
	\label{tab:target_params1}
	\begin{tabular}{lcccc}
    \hline
    & \multicolumn{2}{c}{EBLM J0239-20} 		& \multicolumn{2}{c}{EBLM J0540-17}  	\\
    & \textit{CHEOPS} 			& \textit{TESS} 		& \textit{CHEOPS} 		& \textit{TESS}		\\
    \hline
    Model parameters &&&&\\
    $\rm T_0 (BJD)$		& 2163.70805 $\pm$ 0.00015  & 1413.46145 $\pm$ 0.00012  & 2209.12086 $\pm$0.00021 & 1470.51285 $\pm$ 0.00030\\
    P (days) 			& $\rm 2.778691 (fixed) $   & 2.778691 $\pm$ 0.000001   & 6.004940 (fixed)      & $\rm 6.004940 \pm 0.000003$  \\
    D 					& 0.01679 $\pm$ 0.00019     & 0.016716 $\pm$ 0.000092	    & 0.01404 $\pm$ 0.00021 & 0.01381 $\pm$ 0.00018\\
    W 					& 0.05268 $\pm$ 0.00037     & 0.05286 $\pm$ 0.00015	    & 0.03818 $\pm$ 0.00019 & 0.03827 $\pm$ 0.00018\\
    b 					& 0.654 $\pm$ 0.014	        & 0.6428 $\pm$ 0.0092	        & 0.167 $\pm$ 0.105     & 0.253 $\pm$ 0.089 \\
    $\rm f_c$ 			& 0.0 (fixed)	            & 0.0 (fixed)	            & 0.0 (fixed)           & 0.0 (fixed)\\
    $\rm f_s$ 			& 0.0 (fixed)	            & 0.0 (fixed)	            & 0.0 (fixed)           & 0.0 (fixed)\\
    L 					& $\rm(3.68\pm0.45)\times10^{-4}$	& $\rm(7.30\pm0.42)\times10^{-4}$	& $\rm(3.66\pm0.53)\times10^{-4}$	& $\rm(6.61\pm0.78)\times10^{-4}$\\
    $\rm h_1$ 			& 0.766 $\pm$ 0.020	        & 0.836 $\pm$ 0.011	        & 0.767 $\pm$ 0.015	    & 0.811 $\pm$ 0.013 \\
    $\rm h_2$			& 0.47 $\pm$ 0.22	        & 0.59 $\pm$ 0.20	        & 0.54 $\pm$ 0.18	    & 0.47 $\pm$ 0.21\\
    Derived parameters &&&&  \\
    $\rm R_{2}/R_{1}$ 	& 0.12957 $\pm$ 0.00073	    & 0.12929 $\pm$ 0.00035	    & 0.11850 $\pm$ 0.00087 & 0.11752 $\pm$ 0.00075\\
    $\rm R_{1}/a$ 		& 0.1797 $\pm$ 0.0027	    & 0.1788 $\pm$ 0.0015	    & 0.1084 $\pm$ 0.0018   & 0.1105 $\pm$ 0.0023 \\
    $\rm R_{2}/a$ 		& 0.02288 $\pm$ 0.00042	    & 0.02289 $\pm$ 0.00024	    & 0.01265 $\pm$ 0.00028 & 0.01264 $\pm$ 0.00034\\ 
    $\rm i(^{\circ})$ 	& 83.25 $\pm$ 0.24	        & 83.40 $\pm$ 0.15	        & 88.96 $\pm$ 0.67      & 88.40 $\pm$ 0.59\\
    e 					& 0.0	                    & 0.0	                    & 0.0                   & 0.0 \\
    $\rm \omega(^{\circ})$ 	& --	                    & --	                    & --                    & -- \\
    Absolute parameters &&&& \\
    $\rm a (AU)$            & 0.04106 $\pm$ 0.00076	& 0.04107$\pm$ 0.00076 	& 0.0703 $\pm$     0.0012 	& 0.0703 $\pm$ 0.0012\\
    $\rm R_{2} (R_{\odot})$	& 0.2056 $\pm$   0.0052	& 0.2041 $\pm$ 0.0044	& 0.1939 $\pm$     0.0050	& 0.1959 $\pm$     0.0056\\
    $\rm M_{2} (M_{\odot})$	& 0.1597 $\pm$   0.0059	& 0.1597 $\pm$ 0.0059	& 0.1633 $\pm$     0.0058	& 0.1634 $\pm$     0.0058\\
    $\rm log\,g_{2} (cgs)$ 	& 5.015 $\pm$    0.014	& 5.0214 $\pm$ 0.0076	& 5.075 $\pm$      0.015	& 5.066 $\pm$      0.019\\
    $\rm T_{eff,2} (K)$  	& 3027 $\pm$ 88			& 2982 $\pm$ 71			& 3220 $\pm$ 70				& 3143 $\pm$ 66 \\
    \hline
    \end{tabular}
\end{table*}

\begin{table*}
	\centering
	\caption{The derived parameters for EBLM J0546-18 and EBLM J0719+25 using \textit{CHEOPS} and \textit{TESS} light curve fits with eclipse depths being in the relevant instrumental bandpass.}
	\label{tab:target_params2}
	\begin{tabular}{lcccc}
    \hline
    & \multicolumn{2}{c}{EBLM J0546-18}  & \multicolumn{2}{c}{EBLM J0719+25}\\
    & \textit{CHEOPS} 			& \textit{TESS} 		& \textit{CHEOPS} 		& \textit{TESS}	 \\
    \hline
    Model parameters &&&&\\
    $\rm T_0 (BJD)$		& 2203.71457 $\pm$ 0.00027  & 2174.98660 $\pm$ 0.00032 	& 2216.39007 $\pm$ 0.00024  	& 2559.38262 $\pm$ 0.00019 	  \\
    P (days) 			& 3.191919 (fixed)      	& $\rm 3.191919 \pm 0.000034$ & 7.456295 (fixed) 	& 7.456295 $\pm$ 0.000045		 \\
    D 					& 0.0239 $\pm$ 0.0018 		& 0.02328 $\pm$ 0.00081 	& 0.02145 $\pm$ 0.00051 & 0.02092 $\pm$ 0.00017 		 \\
    W 					& 0.0415 $\pm$ 0.0016		& 0.04020 $\pm$ 0.00047 	& 0.02491 $\pm$ 0.00029	& 0.02456 $\pm$ 0.00018		\\
    b 					& 0.777 $\pm$ 0.040 		& 0.824 $\pm$ 0.013 		& 0.498 $\pm$ 0.033		& 0.520 $\pm$ 0.016		\\
    $\rm f_c$ 			& 0.0 (fixed) 				& 0.0 (fixed) 				& $-$0.2589 $\pm$ 0.0069 	& $-$0.2588 $\pm$ 0.0053 	\\
    $\rm f_s$ 			& 0.0 (fixed) 				& 0.0(fixed) 				& $-$0.116 $\pm$ 0.023	& $-$0.139 $\pm$ 0.022 	\\
    L 					& $\rm(11.0\pm1.3)\times10^{-4}$	& $\rm(17.6\pm1.2)\times10^{-4}$	& $\rm(6.4\pm1.2)\times10^{-4}$	& $\rm(9.32\pm0.65)\times10^{-4}$\\
    $\rm h_1$ 			& 0.44 $\pm$ 0.14$^*$	    & 0.719 $\pm$ 0.100 	    & 0.731 $\pm$ 0.020		& 0.813 $\pm$ 0.013	\\
    $\rm h_2$			& 0.31 $\pm$ 0.14		    & 0.37 $\pm$ 0.24 		    & 0.24 $\pm$ 0.24 		& 0.56 $\pm$ 0.19		\\
    Derived parameters &&&&  \\
    $\rm R_{2}/R_{1}$ 	& 0.1546 $\pm$ 0.0059		& 0.1526 $\pm$ 0.0027 		& 0.1465 $\pm$ 0.0018 	& 0.144625 $\pm$0.000593		\\
    $\rm R_{1}/a$ 		& 0.1533 $\pm$ 0.0057		& 0.1569 $\pm$0.0026 		& 0.0757 $\pm$ 0.0017	& 0.076857 $\pm$ 0.001019		\\
    $\rm R_{2}/a$ 		& 0.0223 $\pm$ 0.0014		& 0.02361 $\pm$0.00034 		& 0.01076 $\pm$ 0.00033 & 0.010941 $\pm$ 0.000176		\\
    $\rm i(^{\circ})$ 	& 83.17 $\pm$ 0.54	    	& 82.58 $\pm$ 0.22 			& 87.84 $\pm$ 0.19		& 87.711 $\pm$ 0.100		\\
    e 					& 0.0						& 0.0 						& 0.0807 $\pm$ 0.0041	& 0.086242 $\pm$ 0.003542		\\
    $\rm \omega(^{\circ})$ 	& -- 						& -- 						& $-$155.9 $\pm$ 4.6		& $-$151.8 $\pm$ 4.3	\\
    Absolute parameters &&&& \\
    $\rm a (AU)$                 & 0.04587 $\pm$0.00080	& 0.04586 $\pm$ 0.00080		& 0.0802 $\pm$   0.0014 & 0.0801 $\pm$     0.0014\\
    $\rm R_{2} (R_{\odot})$	& 0.233 $\pm$  0.013	& 0.2356 $\pm$     0.0072 	& 0.1912 $\pm$ 0.0060 	& 0.1915 $\pm$     0.0044	\\
    $\rm M_{2} (M_{\odot})$	& 0.2129 $\pm$ 0.0075 	& 0.2131 $\pm$     0.0075  	& 0.1584 $\pm$ 0.0056 	& 0.1583 $\pm$     0.0056	\\
    $\rm log\,g_{2} (cgs)$ 	& 5.029 $\pm$  0.047 	& 5.020 $\pm$      0.021 	& 5.075 $\pm$ 0.023	  	& 5.073 $\pm$     0.012	\\
    $\rm T_{eff,2} (K)$  	& 3409 $\pm$ 111 		& 3332 $\pm$ 90 			& 3208 $\pm$ 89 		& 3063 $\pm$ 40			\\
    \hline
    \end{tabular}
    \\
    $^*$ The limb darkening parameters are not well constrained from \textit{CHEOPS} data for EBLM J0546-18 (see discussion in Sec.\,\ref{sec:limb_dark}.)\\
\end{table*}

\begin{table}
	\centering
	\caption{The derived parameters for EBLM J2359+44 using \textit{CHEOPS} and \textit{TESS} light curve fits with eclipse depths being in the relevant instrumental bandpass.}
	\label{tab:target_params3}
	\begin{tabular}{lcc}
    \hline
    & \multicolumn{2}{c}{EBLM J2359+44}  \\
    & \textit{CHEOPS} 		& \textit{TESS}	 \\
    \hline
    Model parameters &&\\
    $\rm T_0 (BJD)$ & 1977.85239 $\pm$ 0.00015  & 1773.4230 $\pm$ 0.0027 \\
    P (days) 		& 11.3627 (fixed)			& $\rm 11.3627 \pm 0.0027$  \\
    D 				& 0.02997 $\pm$ 0.00016		& 0.03015 $\pm$ 0.00023 \\
    W 				& 0.025946 $\pm$ 0.000091	&  0.02611 $\pm$ 0.00017\\
    b 				& 0.096 $\pm$ 0.024			& 0.141 $\pm$ 0.033 \\
    $\rm f_c$ 		& $-$0.05175 $\pm$ 0.00032	&  $-$0.05242 $\pm$ 0.00053\\
    $\rm f_s$ 		& $-$0.68888 $\pm$ 0.00071	& $-$0.68906 $\pm$ 0.00072 \\
    L 				& $\rm(8.91\pm0.63)\times10^{-4}$	& $\rm(20.21\pm0.98)\times10^{-4}$ \\
    $\rm h_1$ 		& 0.7754 $\pm$ 0.0043		& 0.8393 $\pm$ 0.0093 \\
    $\rm h_2$ 		& 0.61 $\pm$ 0.13			& 0.60$\pm$ 0.19 \\
    Derived parameters &&\\
    $\rm R_{2}/R_{1}$ & 0.17311 $\pm$ 0.00045	& 0.17363 $\pm$ 0.00067 \\
    $\rm R_{1}/a$ 	& 0.06971 $\pm$ 0.00033		& 0.07040 $\pm$ 0.00066 \\
    $\rm R_{2}/a$ 	& 0.011990 $\pm$ 0.000077	& 0.01207 $\pm$ 0.00015 \\
    $\rm i(^{\circ})$ 	& 89.619 $\pm$ 0.098	& 89.43 $\pm$ 0.14 \\
    e 				& 0.47724 $\pm$ 0.00098		& 0.47755 $\pm$ 0.00099 \\
    $\rm \omega(^{\circ})$ & $-$94.30 $\pm$ 0.027 		& $-$94.350 $\pm$ 0.044 \\
    Absolute parameters && \\
	$\rm a (AU)$         			& 0.1144 $\pm$ 0.0020 	& 0.1144 $\pm$ 0.0020\\
    $\rm R_{2} (R_{\odot})$  	& 0.2963 $\pm$ 0.0058	& 0.3001 $\pm$ 0.0064 \\
    $\rm M_{2} (M_{\odot})$ 	& 0.293 $\pm$ 0.010		& 0.293 $\pm$ 0.010\\
    $\rm log\,g_{2} (cgs)$   	& 4.9602 $\pm$ 0.0049	& 4.9490 $\pm$ 0.0089 \\
    $\rm T_{eff,2} (K)$ 		& 3465 $\pm$ 46			& 3513 $\pm$ 41 \\
    \hline
    \end{tabular}
\end{table}

\section{Discussion}

We have derived the stellar parameters for both companions for all of our targets thanks to high precision \textit{CHEOPS} light curves. 
%For The primary component, we derived precise masses and surface gravities.
For the M-dwarfs we derive accurate radii with an average uncertainty of $\rm 3.2\pm1.3\%$ and the surface gravity with an average uncertainty of $\rm 0.4\pm0.3\%$. This precision for the surface gravity of M-dwarfs is better then, or hardly reached with state of the art high-resolution spectroscopic measurements of field M-dwarfs  \citep[e.g.][]{Olander21,marfil21}.

\subsection{Radial velocity priors}

We used priors obtained from the radial velocity (RV) orbital parameters eccentricity (e) and longitude of periastron ($\omega$) to fit our \textit{CHEOPS} and \textit{TESS} light curves. Only EBLM J0719+25 and EBLM J2359+44 have eccentricities significantly larger than zero, the others we have fixed to zero eccentricity.
We analysed the effect of imposing RV priors on the \textit{CHEOPS} parameter fit by repeating it with $\rm f_c$ and $\rm f_s$ kept as free parameters. 
Two of our binaries with previously fixed eccentricities, resulted in eccentricties consistent to zero with EBLM J0239-20 (e = 0.028 $\pm$ 0.058) and EBLM J0546-18 (e = 0.0005 $\pm$ 0.0007).
For EBLM J0540-17 and EBLM J0719+25 this fit resulted in a longer MCMC chain, which finally ended with a less uniformly defined PPD for W, which was strongly correlated to $\rm f_c$ and $\rm f_s$. This led to up to 5\% overestimated radii for the M-dwarfs.  
Except for these two stars, the derived model parameters did not deviate more than $1\sigma$ from the parameters listed in Tables\,\ref{tab:target_params1},\,\ref{tab:target_params2}, and\,\ref{tab:target_params3}. 
Nevertheless, we found that for the orbital parameters all resulting uncertainties were about one order of magnitude larger then obtained from the RV fitting alone. We conclude that even for high precision \textit{CHEOPS} light curves, (i) radial velocity measurements are essential to derive precise radii for low mass eclipsing binaries and (ii) our analysis method does not allow to constrain the orbital eccentricity from the light curves better then from radial velocity measurements. 

\subsection{Comparison to \textit{TESS}}

For all targets, we compared our results from \textit{TESS} light curve fitting with the \textit{CHEOPS} results. Both instruments comprise different passbands with the \textit{TESS} having an redder effective wavelength of 745.6\,nm compared to \textit{CHEOPS} with 581.1\,nm\footnote{Filter profiles and effective wavelengths can be accessed using the \href{http://svo2.cab.inta-csic.es/svo/theory/fps3/index.php?}{VSO Filter Profile Service}.}. In this, we do not compare the limb darkening parameters and absolute eclipse depths, since these depend on the instrumental passband. The secondary eclipses are thus 1.5 to 2.5 times deeper in \textit{TESS}, compared to \textit{CHEOPS}. 
%On average the derived parameters agree within $1\sigma$ to the parameters derived with CHEOPS. 
We find a good agreement on the derived radius ratio, inclination and relative primary radii $\rm R_1/a$ (<1\%). As discussed in the previous section, using radial velocity priors is essential to derive precise radii for the M-dwarfs. We find that keeping $\rm f_c$ and $\rm f_s$ as free parameters results in 3-6\% smaller radii for \textit{TESS} light curves (for EBLM J0540-17 and EBLM J0719+25), compared to \textit{CHEOPS}. Using similar radial velocity priors (see chapter \ref{tess_fit}), we find that the derived radii and surface gravity for the M-dwarfs agree well for all targets (on average within 0.9\% and 0.15\% respectively) between \textit{TESS} and \textit{CHEOPS}. 
We find that the uncertainties of the derived parameters from \textit{TESS} light curves are of a similar order, compared to \textit{CHEOPS} results. \textit{TESS} is in favour, for relatively bright secondary companions with deep secondary eclipse and for targets with short orbital periods and thus, many eclipses covered during the monitoring.
%As described in chapter \ref{tess_fit}, we did not impose any priors to the light curve fit, except for keeping zero eccentricities for EBLM J0239-20 and EBLM J0546-18.\ds(add 0540-17?). This leads to some deviation of the fitted eccentricities, as well as slightly smaller radii for both components of \ds{EBLM J0540-17 ($\rm R1/a$ ($\rm -3.7\pm3.8\%$), and $\rm R2/a$ ($-6.7\pm4.1\%$))} and EBLM J0719+25 ($\rm R1/a$ ($\rm -3.2\pm2.3\%$), and $\rm R2/a$ ($-3.1\pm2.5\%$)). This also results is slightly higher surface gravity compared to our CHEOPS parameters. For EBLM J0239-20, EBLM J0546-18, and EBLM J2559+44 
We find that the effective temperature of the M-dwarfs, derived from \textit{TESS} light curves is in agreement with our \textit{CHEOPS} value for EBLM J2359+44, but about 2-4\% cooler for our other targets. We included the result from SW21 for EBLM J1934-42 to analyse for any systematic difference between the effective temperature of the M-dwarf, derived with \textit{TESS} relative to \textit{CHEOPS}. We modelled a constant difference between two instruments using \texttt{EMCEE} to take the RV-jitter of the effective temperatures of both \textit{TESS} and \textit{CHEOPS} into account by weighting the fit by the log-likelihood function. The offset from our sample of six stars results in a slightly lower ($1.11\pm0.99\%$) temperature for \textit{TESS} light curves with a remaining jitter of 0.0076\%.

The small discrepancy in T$\rm _{eff,2}$ might be caused by an underestimation of the secondary eclipse depth (L). In Sec\,\ref{Sec:stel_param} we have discussed that reflected light might lead to an underestimated depth of the secondary eclipse. Nevertheless, this effect affects both passbands of \textit{CHEOPS} and \textit{TESS} in a comparable level and only for the shortest period binaries in our sample. Thus, reflection can not explain this discrepancy. Possible explanations might be uncertainties introduced by the stellar model we used to derive the temperature from the surface brightness, or stellar activity of the primary star, linked to stellar spots which are not accounted for in the eclipse model, we have used.

\subsubsection{Limb darkening parameters}\label{sec:limb_dark}

For our \textit{CHEOPS} and \textit{TESS} fits, we kept the limb darkening parameters $\rm h_1$ and $\rm h_2$ free. To compare our results, we derived expected limb darkening parameters for EBLM J0239-20, EBLM J0540-17, EBLM J0546-18, and EBLM J0719+25 by interpolating the tables for the \textit{TESS} bandpass and Kepler passband (for \textit{CHEOPS} data respectively) published in \cite{maxted18} using the stellar parameters $\rm T_{eff,1}, log\,g_1, and [Fe/H]$ as listed in Table \ref{tab:star_params}, and applying an offset ($\rm h_1 + 0.01$ and $\rm h_2 - 0.045$; \citep{maxted18}). This method did not converge for the hottest star in our sample EBLM J2359+44 since its effective temperature exceeds the tabulated temperature range. Thus, we used the other four targets for this comparison. The expected limb darkening parameters are listed in Table\,\ref{tab:limb_dark}.
We find that $\rm h_1$ agrees on average well with differences of a few percent, while we find larger discrepancies for $\rm h_2$ in the order of several 10 percent similarly in the \textit{CHEOPS} and \textit{TESS} data sets. This finding, as well as the derived uncertainties follow the trend from \cite{maxted18}, (Fig\,4) for $\rm h_2$ to be about one order of magnitude less constrained than $\rm h_1$.
We find some cases of larger uncertainties in \textit{CHEOPS} light curve fits. EBLM J0546-18 we derive about 31\% uncertainty for $\rm h_1$ and the derived parameter, differs more than 70\% from the expectations. This is not surprising, given the large impact parameter which does not allow to constrain the limb darkening parameters for this star. 
%and for EBLM J0719+25 about 99\% uncertainty for $\rm h_2$. In both cases the derived limb darkening parameters, differ more than 70\% from the expectations.
We have repeated the \textit{CHEOPS} and \textit{TESS} fits for these four targets, using the expected limb darkening parameters as priors, but found that introducing these priors will neither improve the fit, nor has it any significant impact on the derived M-dwarf parameters. We, thus, present in Table\,\ref{tab:target_params2} the derived parameters without priors for $\rm h_1$ and $\rm h_2$, noting that the corresponding values are less well constrained with \textit{CHEOPS} compared to \textit{TESS}.

\subsection{Mass--radius diagram}

\begin{figure*}
	\includegraphics[width=\columnwidth]{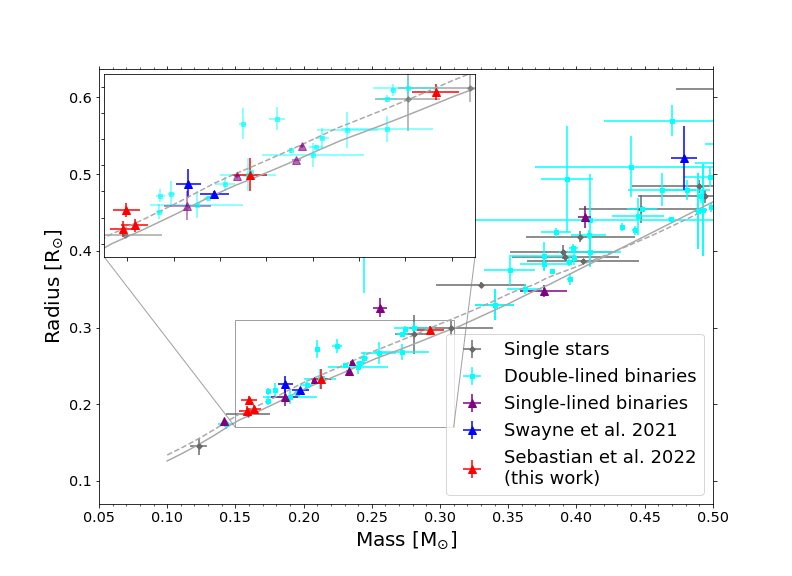}
	\includegraphics[width=\columnwidth]{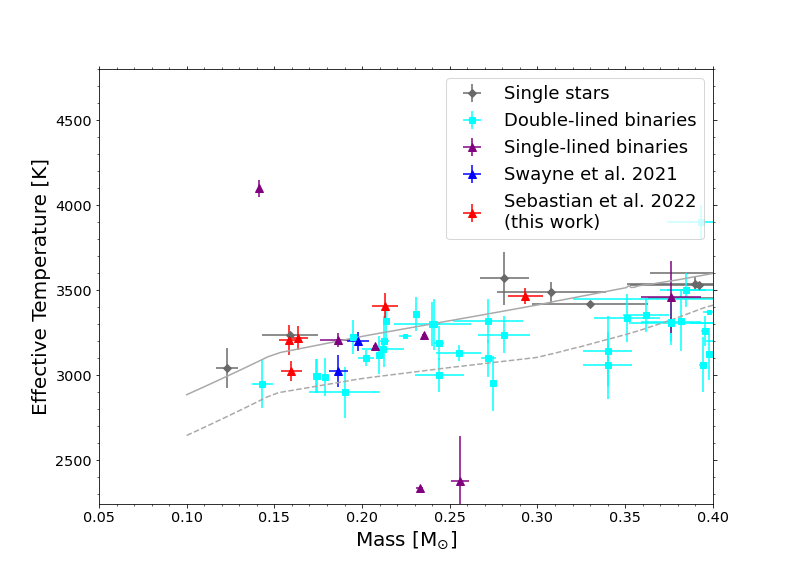}
    \caption{Left: Mass-radius diagram for low mass stars. Triangles: Single lined eclipsing binaries, with \textit{CHEOPS} programme targets highlighted in red and blue.
    Gray, and Cyan squares: single stars and double lined binaries from literature with measured mass, radius, and effective temperature. The zoom in section highlights the MIST model tracks for [Fe/H]=0, grey line, and [Fe/H]=0.25, grey dotted line. Right: Mass-effective temperature diagram of the same data set compared to same MIST models.}
    \label{fig:Mass-Radius}
\end{figure*}

The main goal of the \textit{CHEOPS} programme is to build a well defined mass-radius diagram for stars below the fully convective boundary. In Fig.\,\ref{fig:Mass-Radius} we show our five targets together with the theoretical mass relation from MIST stellar models for 1\,Gyr stars of solar metallicity ([Fe/H]=0.0) as well as for slightly more metal rich stars ([Fe/H]=0.25) \citep{MIST0,MIST1,MIST2}. Similarly to SW21, we compiled a comparison sample of precisely measured low mass stars from literature, classified in single stars, double lined binaries, and single lined binaries. \citep{carter11,nefs13,gillen17,parsons18,smith21,swayne21}. 
We compared the radii with both the MIST and the Exeter/Lyon \citep{Baraffe15} models for solar metallicity. The M-dwarf radius for EBLM J0239-20 is $\rm11.0\pm2.6\%\,(12.5\pm2.6\%$) larger for the MIST (and Exeter/Lyon) model, the others are on average $\rm 2.6\pm1.3\%\,(3.5\pm1.3\%)$ larger compared to both models. Despite most of our targets being within the uncertainties in agreement with the theoretical radii, we observe that they follow the trend of very low mass stars to be slightly larger than predicted by models. In Fig.\,\ref{fig:Mass-Radius} we also show the effective temperature of our five M-dwarfs, the result from SW21, as well as the same literature sample. Our targets effective temperatures follow the overall trend of low mass stars. 
We note that EBLM J0239-20, similarly to EBLM J1934-42 (blue triangle from SW21) have a slightly higher metallicity ($\rm [M/H] > 0.2$). Both stars are slightly larger and cooler, compared to models for stars with solar metallicity. As shown in Fig.\,\ref{fig:Mass-Radius} this trend is predicted by the MIST models for more metal rich stars. But also in this case, both stars are slightly larger than predicted by models for higher metallicity stars.
Fig.\,\ref{fig:Mass-Radius} shows three single lined stars from literature with measured M-dwarf effective temperatures being outliers of more than 500\,K compared to model predictions. These are KIC 1571511B \citep{ofir12} as well as SAO 106989 and HD 24465 \citep{Chaturvedi18}. Populating the low-mass main-sequence with M-dwarfs having precise effective temperature measurements will help us to constrain possible trends for low-mass dwarfs. This is one of the main goals of our \textit{CHEOPS} programme.

Magnetic activity of the primary star, like spot crossing is not accounted for in our eclipse model, thus, can affect the size determinations of the M-dwarfs. We used the \textit{TESS} light curves to search for variability linked to magnetic activity, like rotational pattern and flares. No flares have been found in the \textit{TESS} data set.
EBLM J0239-20 shows a variable modulation of 2-3\% close to the orbital period, most probably linked to stellar activity aligned with the rotational period of the G-dwarf. All our other targets show no or small variability of less than 1\,\%. 
Since we found a good agreement between the M-dwarf radii in the different passbands of \textit{TESS} and \textit{CHEOPS}, we conclude that stellar activity can only have a minor ($\rm <1\%$) effect on the derived M-dwarf radius for the five stars, analysed in this work.
Depending on the actual contrast between the primary star and the M-dwarf the contribution of the M-dwarf is between 300 and 1200\,ppm in \textit{CHEOPS} data. From this we can exclude large flares with exceed relative intensities of 25 to 100\,\% compared to the M-dwarfs average brightness. M-dwarfs with such flaring activity exist but account only for about 10\,\% of the flaring M-dwarfs found in \textit{TESS} \citep{Guenther20}. We can assume that the M-dwarf rotation period is synchronised with the orbital period, since the tidal synchronisation timescale for EBLM systems is about 1\,Gyr or less \citep{barker20}. Thus the M-dwarfs are expected to be fast rotators ($\rm P \lesssim 10\,d$), which are expected to show enhanced activity levels \citep[e.g.][]{morales10,wright18}. Activity induced photometric variations, observed for field M-dwarfs is typically in the order of 1\,\% of the M-dwarfs average brightness \citep{medina20}. This results in an expected photometric variability in the order of 10\,ppm for active M-dwarfs which is below the detection efficiency in our data.

Reflected light from the primary star (See discussion in Sec\,\ref{Sec:stel_param}) can cause an underestimated radius of the M-dwarfs. We note that this effect is negligible for the five binaries analysed in this work, as it would result in a relative underestimation of about 100\,ppm of the M-dwarfs radius for the shortest period binaries in our sample. 

%wright18 - flate activity-rotation curve
%morales10 activity of low-mass stars
\section{Summary}

Within the framework of our EBLM project, we initiated a \textit{CHEOPS} observing programme of 23 low-mass stars to measure precise stellar parameters as well as effective temperatures. In this paper, we have analysed high precision \textit{CHEOPS} light curves of primary and secondary eclipses for five eclipsing binaries with low mass companions. 
Using the \texttt{qpower2} transit model, of \texttt{PYCHEOPS}, we find an average uncertainty of $\rm 3.2\pm1.3\%$ for the M-dwarfs radius and $\rm 0.4\pm0.3\%$ for the M-dwarfs surface gravity. Thus, using precision light curves allowed us to overcame the larger uncertainties to derive stellar parameters typically involved with high-resolution spectroscopy. We have derived the M-dwarfs effective temperature from the contrast between primary and secondary eclipses and the metallicity from spectroscopic analysis of the primary star, assuming equal metallicities of both components.

This allows us to compare the M-dwarfs parameters to theoretical structural models, like the MIST models. We find that all our M-dwarfs are on average larger, but agree within the uncertainty with the model predictions. This is also true for low-mass M-dwarfs with enhanced metallicity, which follow the predicted trend of having a larger radius as well as a cooler effective temperature. 
Up to now, the stellar models, as well as our transit model do not include stellar activity. We have analysed \textit{TESS} light curves for all our five targets and find a good (better than 1\%) agreement on the M-dwarf radius in the different passband of both instruments.
Given the absence of strong activity indicated variability and flare activity as well as this good agreement, we conclude that stellar activity does not play a strong role in the derived uncertainties for our five stars. This result is of particular importance for more active stars on our \textit{CHEOPS} programme, where activity induced changes in parameters between the \textit{TESS} and \textit{CHEOPS} passbands might need to be accounted for.
%makes those targets with combined analysis in the \textit{TESS} and \textit{CHEOPS} passband particular important, since activity induced parameter changes might need to be accounted for in more active stars in the \textit{CHEOPS} programme.
We have analysed the dependence of derived M-dwarf parameters with priors used in the fit. We find that limb darkening parameters as well as orbital parameters like the eccentricity and the argument of periastron are not well constrained from our model fit. Nevertheless, we find that, other than the limb darkening coefficients, precise orbital parameters, obtained from radial velocity observations are crucial to derive M-dwarf radii better than 5\%.

Together with SW21, we increased the sample to eight low-mass stars, with precise measured radii from \textit{CHEOPS} data. Due to the fact that the F,G,K-type primary companions are single lined binaries, that allow high-precision orbital characterisation as well as the determination of precise stellar parameters like metallicity, this survey, once completed, will allow us to  empirically shed light on the radius inflation problem for very low mass stars.

\section*{Acknowledgements}

\textit{CHEOPS} is an ESA mission in partnership with Switzerland with important contributions to the payload and the ground segment from Austria, Belgium, France, Germany, Hungary, Italy, Portugal, Spain, Sweden, and the United Kingdom. The \textit{CHEOPS} Consortium would like to gratefully acknowledge the support received by all the agencies, offices, universities, and industries involved. Their flexibility and willingness to explore new approaches were essential to the success of this mission. Spectroscopic data were obtained via observing time allocations at OHP awarded by the French PNP (18B.PNP.SAN1, 19A.PNP.SANT). Some of the observations reported in this paper were obtained with the Southern African Large Telescope (SALT). This research has made use of the services of the ESO Science Archive Facility. This research is also supported work funded from the European Research Council (ERC) the European Union’s Horizon 2020 research and innovation programme (grant agreement n◦803193/BEBOP). 
PM acknowledges support from STFC research grant number ST/M001040/1. MIS acknowledges support from STFC grant number ST/T506175/1
S.G.S. acknowledge support from FCT through FCT contract nr. CEECIND/00826/2018 and POPH/FSE (EC). SH gratefully acknowledges CNES funding through the grant 837319. YA and MJH acknowledge the support of the Swiss National Fund under grant 200020\_172746. 
We acknowledge support from the Spanish Ministry of Science and Innovation and the European Regional Development Fund through grants ESP2016-80435-C2-1-R, ESP2016-80435-C2-2-R, PGC2018-098153-B-C33, PGC2018-098153-B-C31, ESP2017-87676-C5-1-R, MDM-2017-0737 Unidad de Excelencia Maria de Maeztu-Centro de Astrobiologí­a (INTA-CSIC), as well as the support of the Generalitat de Catalunya/CERCA programme. The MOC activities have been supported by the ESA contract No. 4000124370. 
S.C.C.B. acknowledges support from FCT through FCT contracts nr. IF/01312/2014/CP1215/CT0004. 
XB, SC, DG, MF and JL acknowledge their role as ESA-appointed \textit{CHEOPS} science team members. 
ABr was supported by the SNSA. 
ACC acknowledges support from STFC consolidated grant numbers ST/R000824/1 and ST/V000861/1, and UKSA grant number ST/R003203/1. 
This project was supported by the CNES. 
The Belgian participation to \textit{CHEOPS} has been supported by the Belgian Federal Science Policy Office (BELSPO) in the framework of the PRODEX Program, and by the University of Liège through an ARC grant for Concerted Research Actions financed by the Wallonia-Brussels Federation. 
L.D. is an F.R.S.-FNRS Postdoctoral Researcher. 
This work was supported by FCT - Fundação para a Ciência e a Tecnologia through national funds and by FEDER through COMPETE2020 - Programa Operacional Competitividade e Internacionalizacão by these grants: UID/FIS/04434/2019, UIDB/04434/2020, UIDP/04434/2020, PTDC/FIS-AST/32113/2017 \& POCI-01-0145-FEDER- 032113, PTDC/FIS-AST/28953/2017 \& POCI-01-0145-FEDER-028953, PTDC/FIS-AST/28987/2017 \& POCI-01-0145-FEDER-028987, O.D.S.D. is supported in the form of work contract (DL 57/2016/CP1364/CT0004) funded by national funds through FCT. 
B.-O.D. acknowledges support from the Swiss National Science Foundation (PP00P2-190080). 
This project has received funding from the European Research Council (ERC) under the European Union’s Horizon 2020 research and innovation programme (project {\sc Four Aces}. 
grant agreement No 724427). It has also been carried out in the frame of the National Centre for Competence in Research PlanetS supported by the Swiss National Science Foundation (SNSF). DE acknowledges financial support from the Swiss National Science Foundation for project 200021\_200726. 
MF and CMP gratefully acknowledge the support of the Swedish National Space Agency (DNR 65/19, 174/18). 
DG gratefully acknowledges financial support from the CRT foundation under Grant No. 2018.2323 ``Gaseousor rocky? Unveiling the nature of small worlds''. 
M.G. is an F.R.S.-FNRS Senior Research Associate. 
KGI is the ESA \textit{CHEOPS} Project Scientist and is responsible for the ESA \textit{CHEOPS} Guest Observers Programme. She does not participate in, or contribute to, the definition of the Guaranteed Time Programme of the \textit{CHEOPS} mission through which observations described in this paper have been taken, nor to any aspect of target selection for the programme. 
This work was granted access to the HPC resources of MesoPSL financed by the Region Ile de France and the project Equip@Meso (reference ANR-10-EQPX-29-01) of the programme Investissements d'Avenir supervised by the Agence Nationale pour la Recherche. 
ML acknowledges support of the Swiss National Science Foundation under grant number PCEFP2\_194576. 
LBo, GBr, VNa, IPa, GPi, RRa, GSc, VSi, and TZi acknowledge support from CHEOPS ASI-INAF agreement n. 2019-29-HH.0. 
This work was also partially supported by a grant from the Simons Foundation (PI Queloz, grant number 327127). 
IRI acknowledges support from the Spanish Ministry of Science and Innovation and the European Regional Development Fund through grant PGC2018-098153-B- C33, as well as the support of the Generalitat de Catalunya/CERCA programme. 
GyMSz acknowledges the support of the Hungarian National Research, Development and Innovation Office (NKFIH) grant K-125015, a PRODEX Institute Agreement between the ELTE E\"otv\"os Lor\'and University and the European Space Agency (ESA-D/SCI-LE-2021-0025), the Lend\"ulet LP2018-7/2021 grant of the Hungarian Academy of Science and the support of the city of Szombathely. 
V.V.G. is an F.R.S-FNRS Research Associate. 
NAW acknowledges UKSA grant ST/R004838/1. V.K. acknowledges support from NSF award AST2009501. This work was also supported by the STFC PATT Travel grant ST/S001301/1.
We thank the reviewer for valuable comments and suggestions.

%%%%%%%%%%%%%%%%%%%%%%%%%%%%%%%%%%%%%%%%%%%%%%%%%%
\section*{Data Availability}

All \textit{CHEOPS} data and data products are publicly available via the \href{http://dace.unige.ch/}{Data Analysis Center for
Exoplanets} web platform. %PFLM after the one year guaranteed time period.
This paper includes data collected by the \textit{TESS} mission, which
is publicly available from the Mikulski Archive for Space Telescopes (MAST) at the Space Telescope Science Institute (STScI)
(\url{https://mast.stsci.edu}). Funding for the \textit{TESS} mission is provided
by the NASA Explorer Program directorate. STScI is operated by the
Association of Universities for Research in Astronomy, Inc., under
NASA contract NAS 5-26555. We acknowledge the use of public
\textit{TESS} Alert data from pipelines at the \textit{TESS} Science Office and at
the \textit{TESS} Science Processing Operations Center.
SOPHIE high-resolution spectra are available trough the data archives of the Observatoire de Haute-Provence via \url{http://atlas.obs-hp.fr/}. Programme ID were 18B.PNP.SAN1, and 19A.PNP.SANT.



\bibliographystyle{mnras}
\bibliography{library} % if your bibtex file is called example.bib

%\begin{thebibliography}{99}
%\end{thebibliography}

%%%%%%%%%%%%%%%%%%%%%%%%%%%%%%%%%%%%%%%%%%%%%%%%%%

%%%%%%%%%%%%%%%%% APPENDICES %%%%%%%%%%%%%%%%%%%%%

\appendix

\section{Decorrelation parameters fitted from CHEOPS fits}

\begin{table*}[!h]
	\centering
	\caption{Decorrelation parameters fitted from \textit{CHEOPS} multivisit analysis for each visit (in the same order as in Table~\ref{tab:CHEOPS_log}). The parameters are: Image background level (dfdbg), PSF centroid position (dfdx and dfdy), time (dfdt), and aperture contamination (dfdcontam).}
	\label{tab:decorr}
	\begin{tabular}{lccccccc}
    \hline
    Target & Eclipse            & dfdbg           & dfdx          & dfdy           & dfdt                 & dfdcontam   \\
           &                    & ($\rm 10^{-3}$) &($\rm 10^{-4}$)& ($\rm 10^{-3}$)& ($\rm 10^{-2}d^(-1)$)& ($\rm 10^{-3}$) \\\hline
    EBLM J0239-20   & primary & -- & -- & -- & -- & --     \\
                    & secondary & $\rm1.57 \pm 0.90$ & -- & $\rm0.311 \pm 0.085$ & $\rm2.924 \pm 0.029$ & --     \\
                    & secondary & $\rm1.21 \pm 0.23$ & -- & -- & $\rm1.680 \pm 0.029$ & --    \\
    EBLM J0540-17   & primary & $\rm1.20 \pm 0.82$ & $\rm7.33 \pm 1.79$ & -- & $\rm-0.31 \pm 0.43$ & --     \\
                    & secondary & $\rm0.71 \pm 0.77$ & -- & -- & -- & --     \\
                    & secondary & -- & -- & $\rm-0.51 \pm 0.14$ & $\rm0.163 \pm 0.036$ & --     \\
                    & secondary & -- & $\rm5.95 \pm 1.71$ & $\rm-0.87 \pm 0.17$ & -- & --     \\
                
    EBLM J0546-18   & primary & $\rm4.80 \pm 0.87$ & -- & $\rm0.78 \pm 0.23$ & -- & $\rm-1.73 \pm 0.56$     \\
                    & secondary & -- & -- & -- & -- & -1.59 +/- 0.83     \\
                    & secondary & $\rm2.85 \pm 0.66$ & $\rm11.32 \pm 2.51$ & -- & $\rm1.367 \pm 0.079$ & --     \\
                    
    EBLM J0719+25   & primary & -- & -- & -- & $\rm-0.496 \pm 0.060$ & --     \\
                    & secondary & $\rm1.22 \pm 0.93$ & -- & -- & $\rm0.291 \pm 0.061$ & --     \\
                    & secondary & -- & -- & -- & -- & --     \\
                    
    EBLM J2359+44   & secondary & $\rm0.83 \pm 0.40$ & -- & $\rm0.208 \pm 0.088$ & -- & $\rm-0.48 \pm 0.27$     \\
                    & primary & $\rm0.83 \pm 0.26$ & -- & -- & -- & --     \\
    \hline
    \end{tabular}
    \\
    
\end{table*}

\section{Radial velocity measurements}

\begin{table}
	\centering
	\caption{Radial velocity measurements for EBLM J0719+25}
	\label{tab:rv_J0719+25}
	\begin{tabular}{lccc}
    \hline
    BJD - 2400000 & RV [$\rm km\,s^{-1}$] & RV error [$\rm km\,s^{-1}$] & Source \\\hline
    58436.57258 & $-$5.9492 & 0.0079 & SOPHIE\\
    58438.59676 & 12.5703 & 0.0057 & SOPHIE\\
    58536.40291 & 11.1258 & 0.0058 & SOPHIE\\
    58538.42658 & $-$9.091 & 0.012 & SOPHIE\\
    58542.39085 & 10.1391 & 0.0047 & SOPHIE\\
    58562.39379 & $-$15.9404 & 0.0073 & SOPHIE\\
    58566.37826 & 10.2797 & 0.0053 & SOPHIE\\
    58761.63689 & $-$3.306 & 0.011 & SOPHIE\\
    \hline
    \end{tabular}
    \\
    
\end{table}

\begin{table}
	\centering
	\caption{Radial velocity measurements for EBLM J2359+44}
	\label{tab:rv_J2359+44}
	\begin{tabular}{lccc}
    \hline
    BJD - 2400000 & RV [$\rm km\,s^{-1}$] & RV error [$\rm km\,s^{-1}$] & Source \\\hline
    53310.6391  & $-$19.07   & 0.42   & Poleski et al.\\
    53311.7990  & $-$26.36   & 0.50   & Poleski et al. \\
    58436.31776 & $-$33.537 & 0.011 & SOPHIE\\
    58438.40839 & 2.8147   & 0.0086 & SOPHIE\\
    58685.56693 & $-$29.4759 & 0.012 & SOPHIE\\
    58704.54724 & $-$8.063  & 0.014 & SOPHIE\\
    58729.61888 & $-$20.846 & 0.013 & SOPHIE\\
    58734.5406  & 11.81   & 0.015 & SOPHIE\\
    58754.47118 & $-$33.987 & 0.015 & SOPHIE\\
    58765.46162 & $-$31.893 & 0.011 & SOPHIE\\
    59030.57795 & 10.110  & 0.011 & SOPHIE\\
    59043.50347 & 1.726    & 0.014 & SOPHIE\\
    59045.53151 & $-$9.040  & 0.012 & SOPHIE\\
    59071.56389 & $-$27.920 & 0.012 & SOPHIE\\
    59077.5554  & 1.898   & 0.012 & SOPHIE\\
    59094.51791 & $-$29.440 & 0.011 & SOPHIE\\
    59100.57485 & 0.226    & 0.012 & SOPHIE\\
    \hline
    \end{tabular}
    \\
\end{table}

\section{Expected limb darkening coefficients}

\begin{table}
	\centering
	\caption{Expected limb darkening coefficients derived for \textit{TESS} and \textit{CHEOPS} passbands.}
	\label{tab:limb_dark}
	\begin{tabular}{lcccc}
    \hline
    Target & \multicolumn{2}{c}{\textit{CHEOPS}} & \multicolumn{2}{c}{\textit{TESS}}\\
                    & $\rm h_1$ & $\rm h_2$ & $\rm h_1$ & $\rm h_2$\\\hline
    EBLM J0239-20   & 0.743$\pm$0.012 & 0.40$\pm$0.05 & 0.798$\pm$0.012 & 0.39$\pm$0.05 \\
    EBLM J0540-17   & 0.773$\pm$0.011 & 0.41$\pm$0.05 & 0.826$\pm$0.011 & 0.38$\pm$0.05 \\
    EBLM J0546-18   & 0.771$\pm$0.011 & 0.41$\pm$0.05 & 0.822$\pm$0.011 & 0.37$\pm$0.05 \\
    EBLM J0719+25   & 0.754$\pm$0.011 & 0.41$\pm$0.05 & 0.808$\pm$0.011 & 0.39$\pm$0.05 \\
    \hline
    \end{tabular}
    \\
\end{table}

\newpage
\section{\textit{TESS} fits}

\begin{figure*}
	\includegraphics[width=\linewidth]{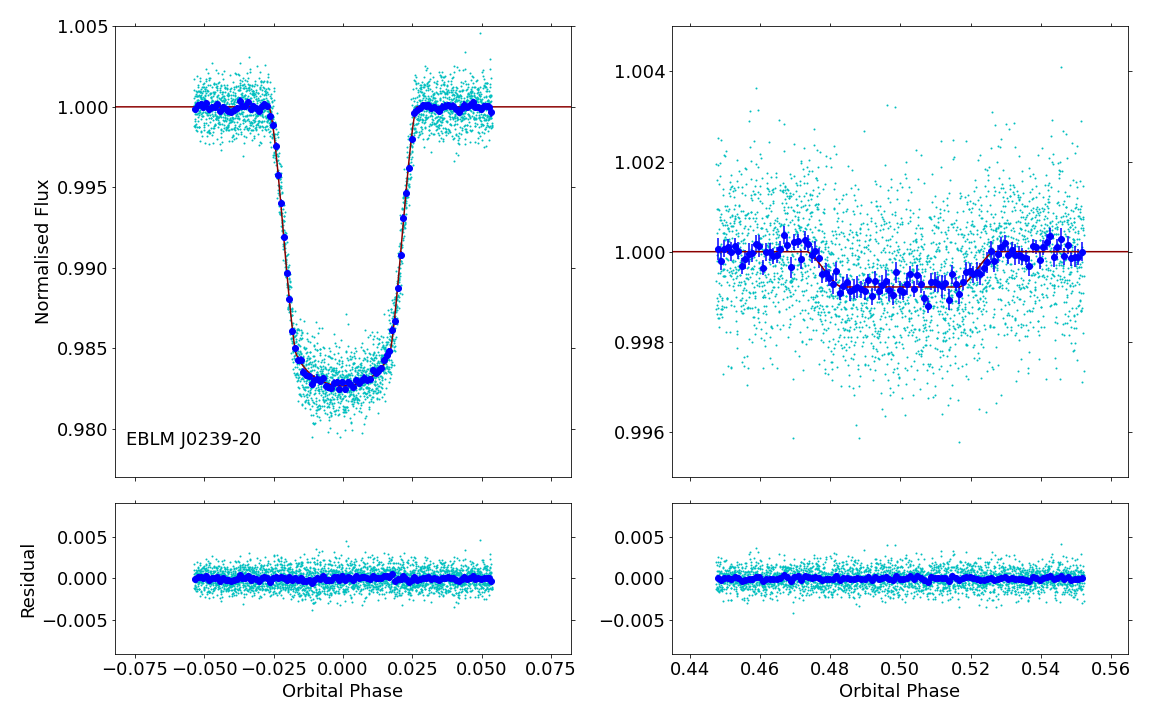}
	\includegraphics[width=\linewidth]{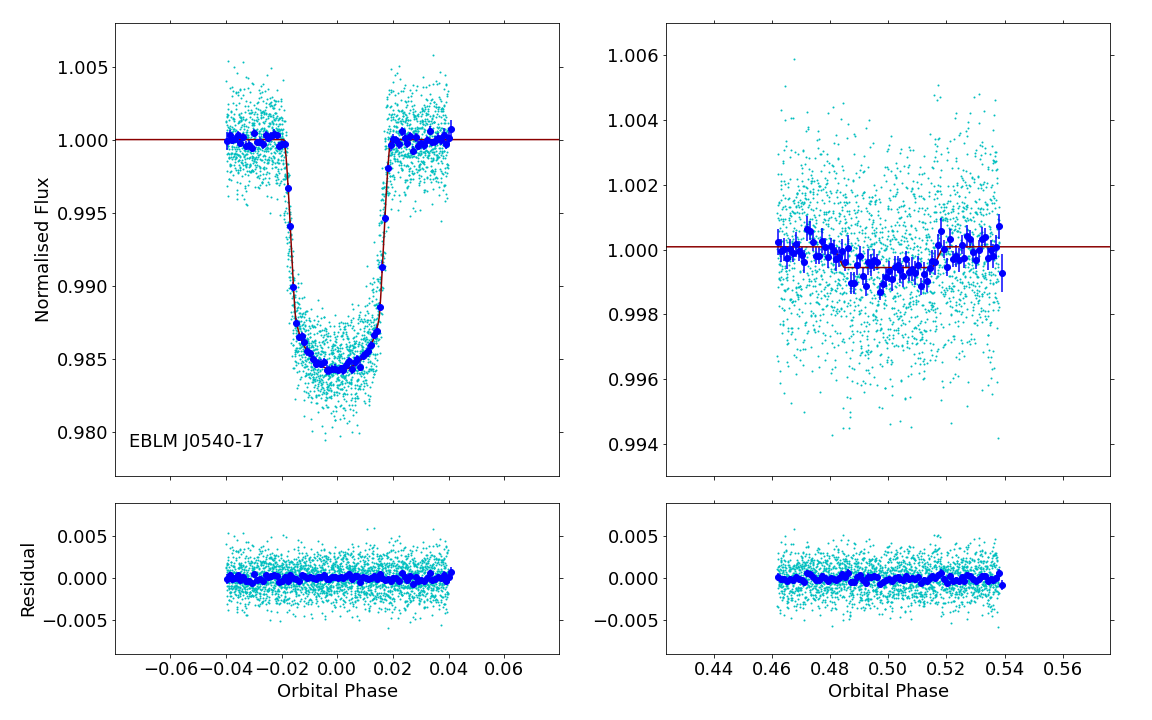}

    \caption{Fitted \textit{TESS} light curves of all targets in phase intervals around the primary and secondary eclipse events. The observed
data points are shown in cyan. The fitted light curve is shown in red. The residual of the fit is displayed the fitted curves.}
    \label{fig:Tess_lcs1}
\end{figure*}

\begin{figure*}
	%\includegraphics[width=\linewidth]{Images/0546lc_tessdiffDR.png}
	\includegraphics[width=\linewidth]{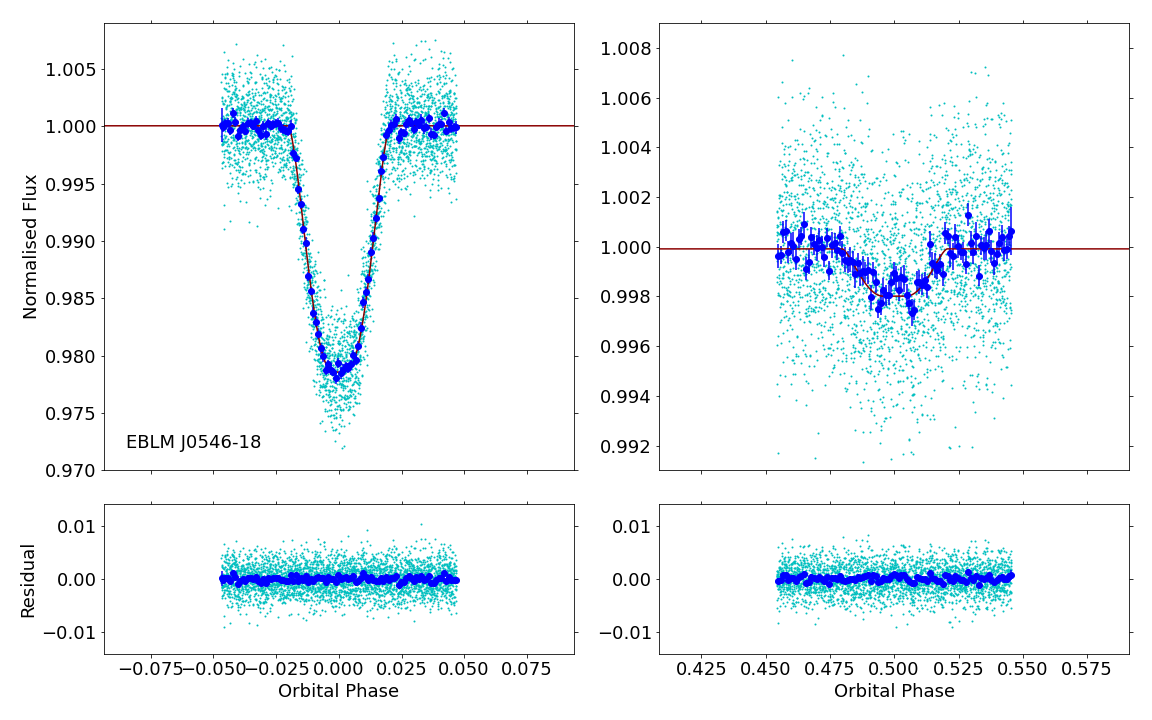}
	\includegraphics[width=\linewidth]{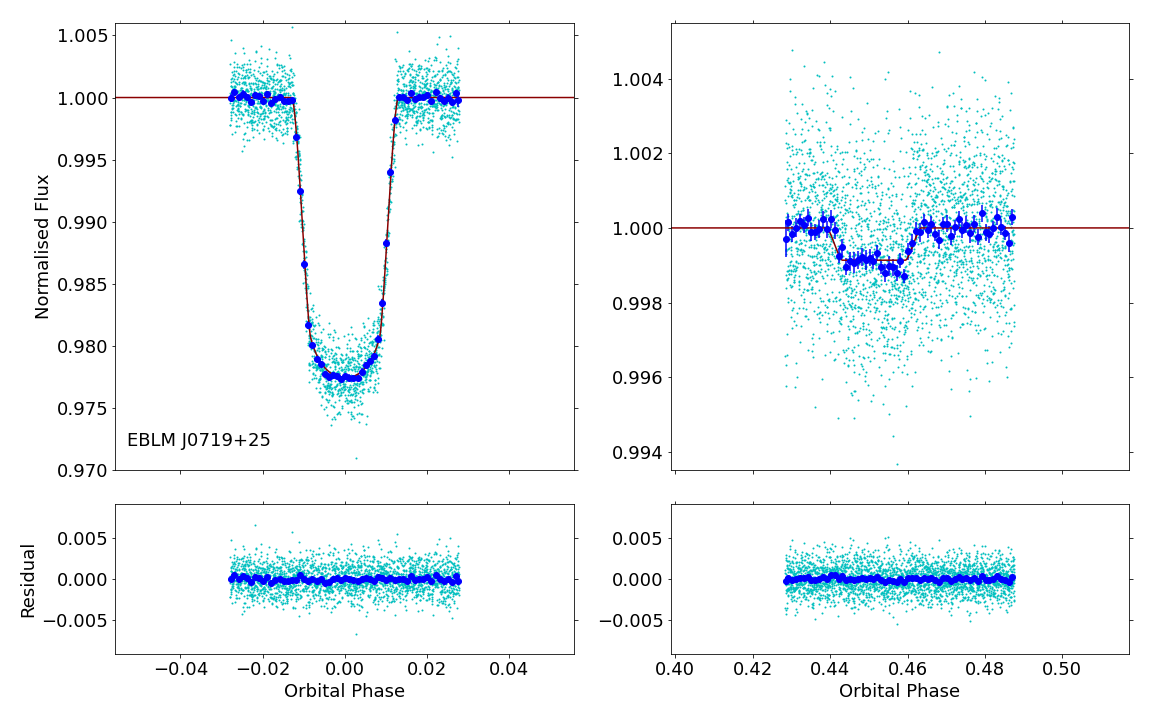}
    
    \caption{Fitted \textit{TESS} light curves of all targets in phase intervals around the primary and secondary eclipse events. The observed
data points are shown in cyan. The fitted light curve is shown in red. The residual of the fit is displayed the fitted curves.}
    \label{fig:Tess_lcs2}
\end{figure*}

\begin{figure*}
	\includegraphics[width=\linewidth]{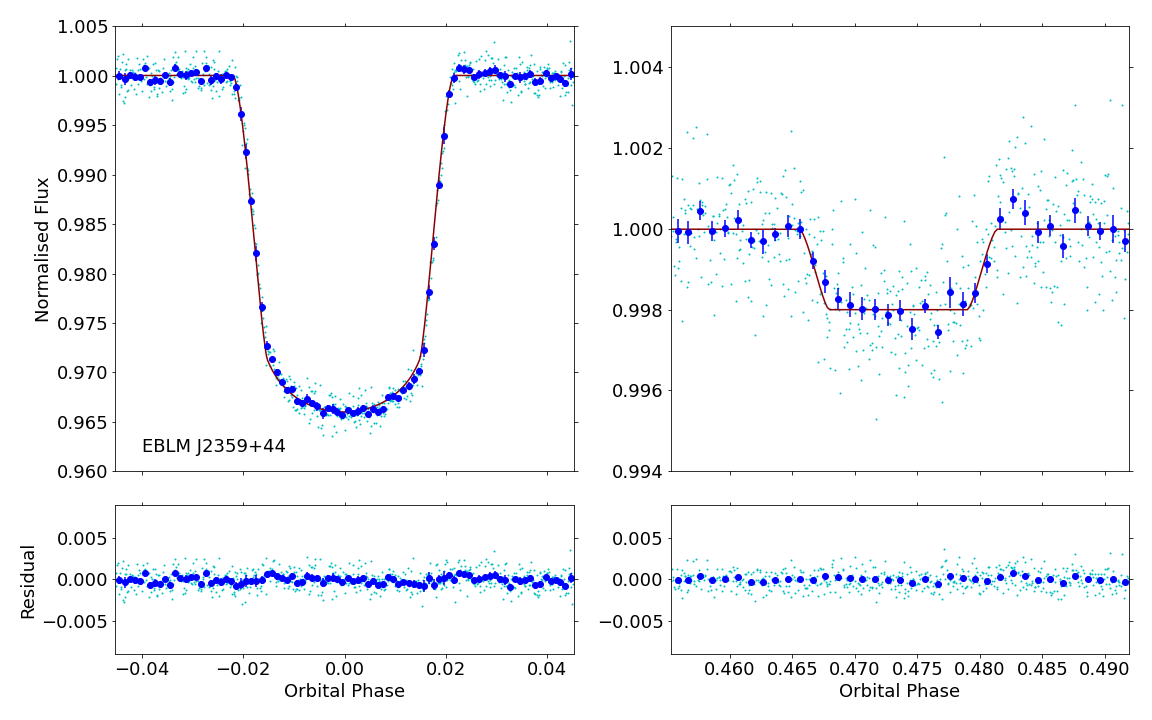}

    \caption{Fitted \textit{TESS} light curves of all targets in phase intervals around the primary and secondary eclipse events. The observed
data points are shown in cyan. The fitted light curve is shown in red. The residual of the fit is displayed the fitted curves.}
    \label{fig:Tess_lcs1}
\end{figure*}

\section{\textit{CHEOPS} fits}

\begin{figure*}
	
	\includegraphics[width=\linewidth]{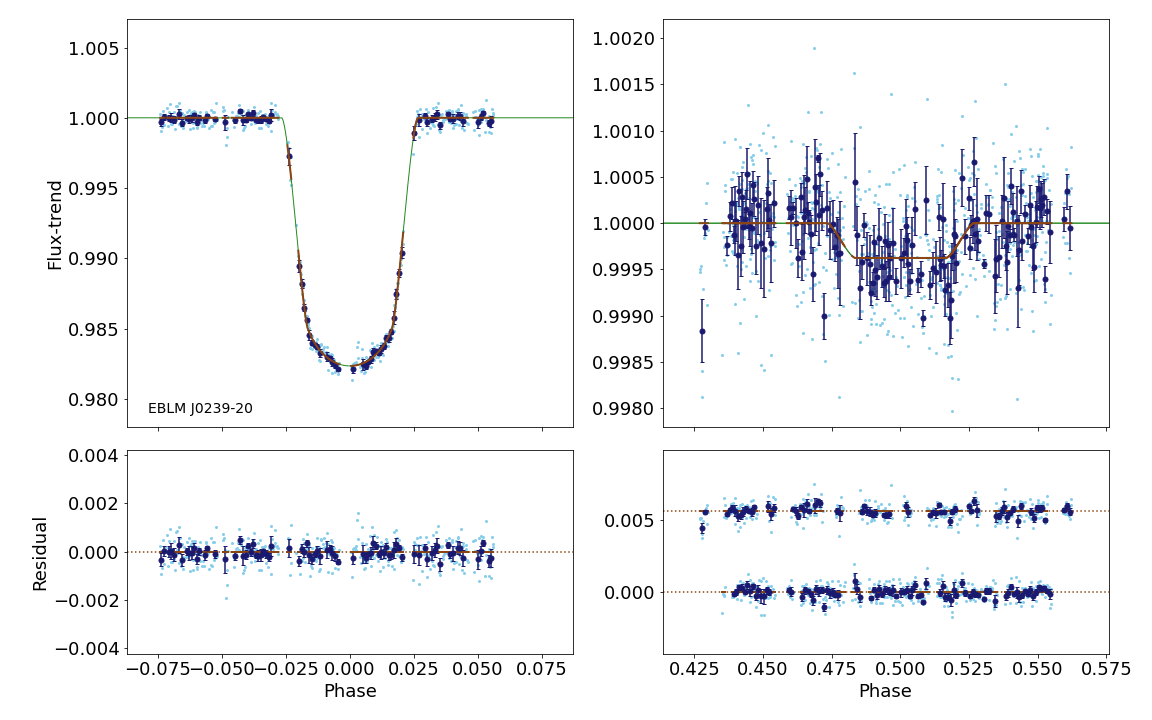}
	\includegraphics[width=\linewidth]{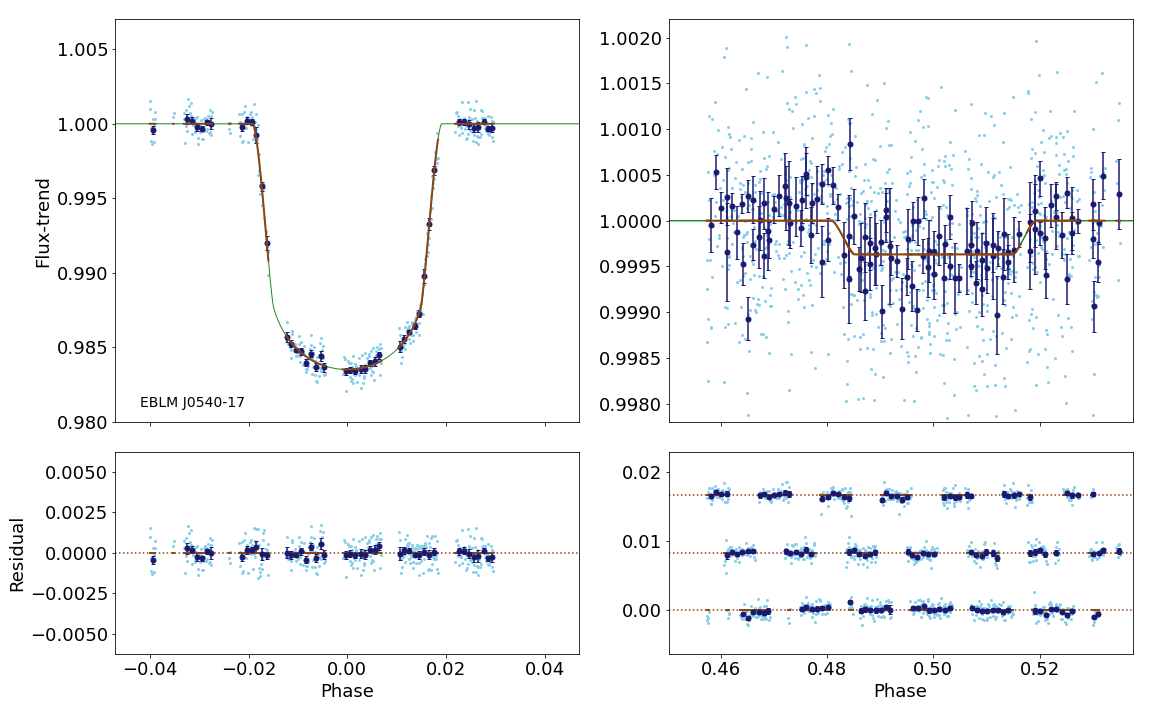}
    \caption{Fitted \textit{CHEOPS} light curves of all targets in phase intervals around the primary and secondary eclipse events. The observed
data points are shown in cyan. The fitted light curve is shown in red. The residual of the fit is displayed in blue below the fitted curves.}
    \label{fig:CHEOPS_lcs1}
\end{figure*}

\begin{figure*}
	\includegraphics[width=\linewidth]{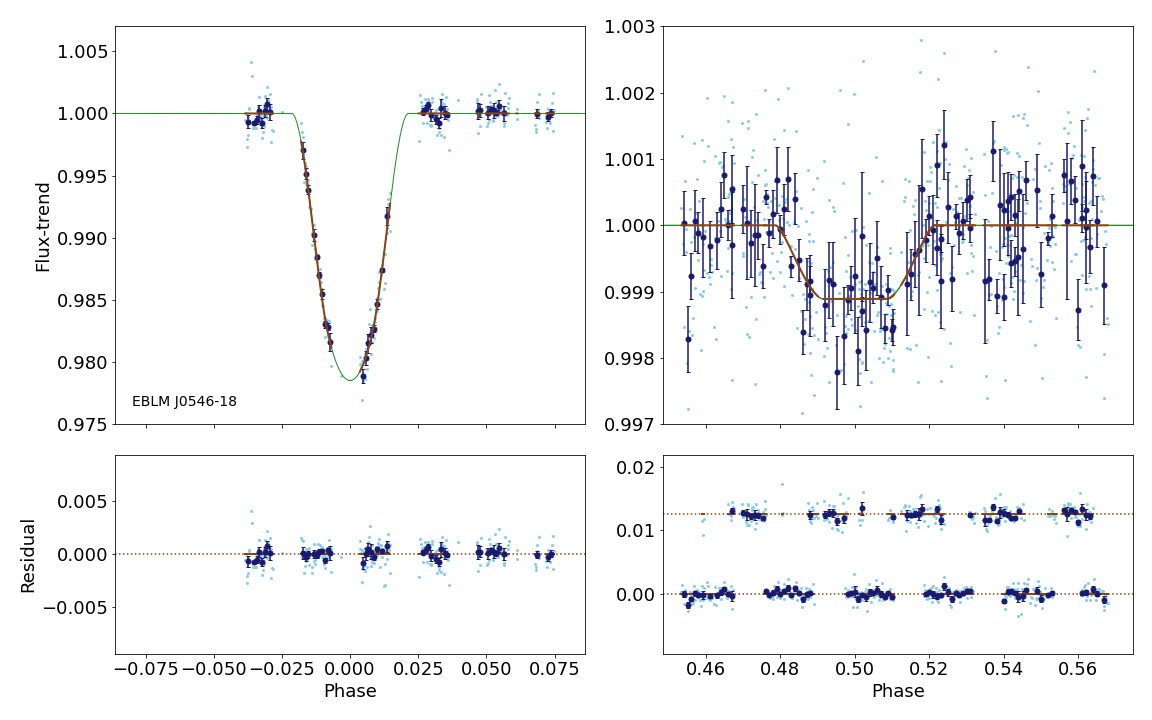}
	\includegraphics[width=\linewidth]{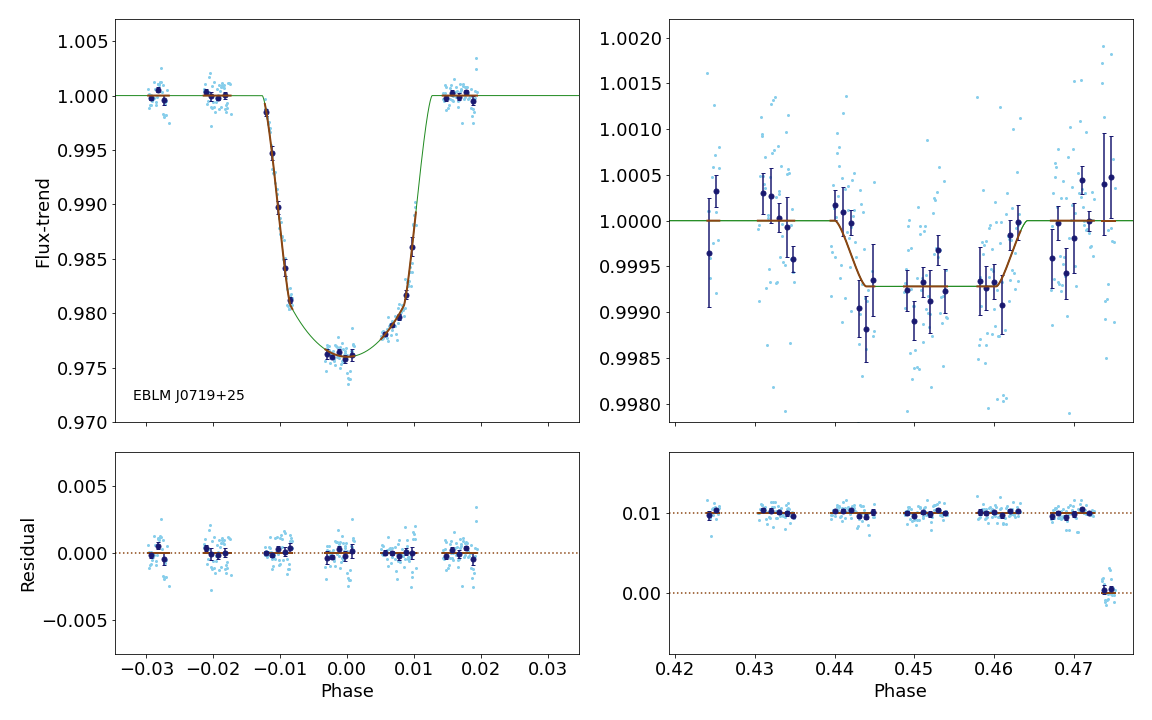}
    \caption{Fitted \textit{CHEOPS} light curves of all targets in phase intervals around the primary and secondary eclipse events. The observed
data points are shown in cyan. The fitted light curve is shown in red. The residual of the fit is displayed in blue below the fitted curves.}
    \label{fig:CHEOPS_lcs2}
\end{figure*}

\begin{figure*}
	\includegraphics[width=\linewidth]{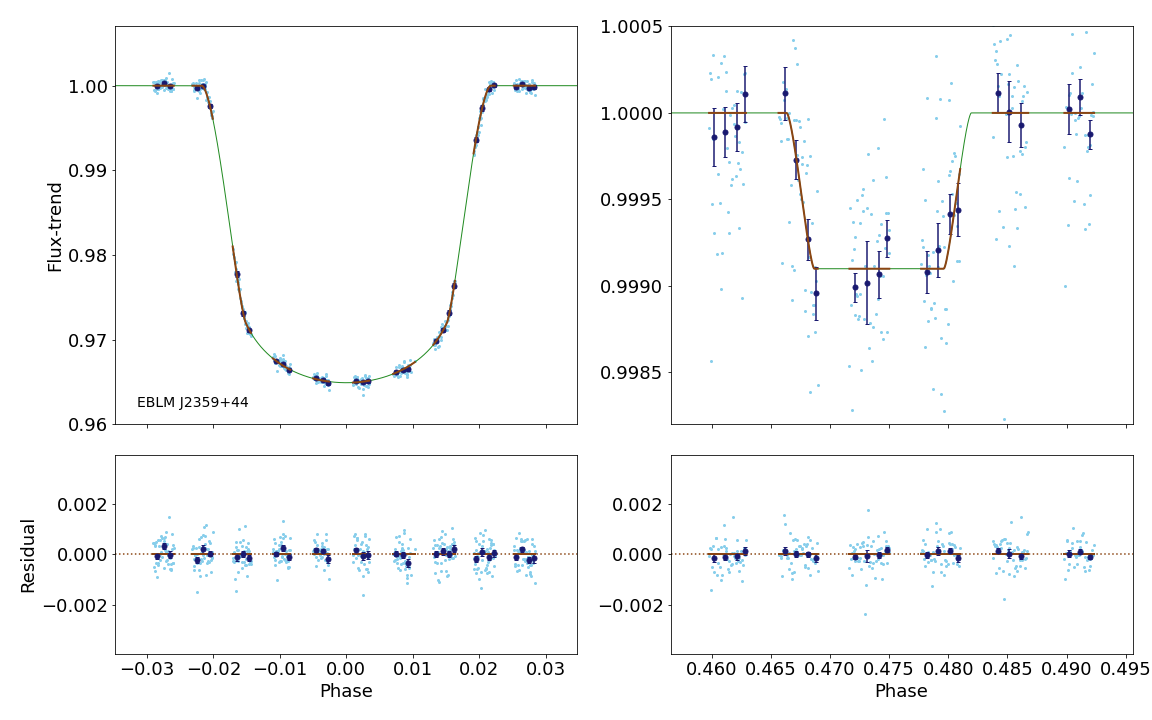}
    \caption{Fitted \textit{CHEOPS} light curves of all targets in phase intervals around the primary and secondary eclipse events. The observed
data points are shown in cyan. The fitted light curve is shown in red. The residual of the fit is displayed in blue below the fitted curves.}
    \label{fig:CHEOPS_lcs3}
\end{figure*}

\onecolumn
\clearpage
\section*{Affiliations}

$^{1}$ School of Physics \& Astronomy, University of Birmingham, Edgbaston, Birmimgham, B15 2TT, UK \\
$^{2}$ Astrophysics Group, Keele University, Staffordshire, ST5 5BG, United Kingdom \\
$^{3}$ Instituto de Astrofisica e Ciencias do Espaco, Universidade do Porto, CAUP, Rua das Estrelas, 4150-762 Porto, Portugal \\
$^{4}$ Department of Astronomy, Stockholm University, AlbaNova University Center, 10691 Stockholm, Sweden \\
$^{5}$ Observatoire Astronomique de l'Université de Genève, Chemin Pegasi 51, Versoix, Switzerland \\
$^{6}$ Aix Marseille Univ, CNRS, CNES, LAM, 38 rue Frédéric Joliot-Curie, 13388 Marseille, France \\
$^{7}$ Department of Physics, University of Warwick, Gibbet Hill Road, Coventry CV4 7AL, United Kingdom \\
$^{8}$ Department of Astronomy, The Ohio State University, 4055 McPherson Laboratory, Columbus, OH 43210, USA \\
$^{9}$ Department of Space, Earth and Environment, Chalmers University of Technology, Onsala Space Observatory, 439 92 Onsala, Sweden \\
$^{10}$ Physikalisches Institut, University of Bern, Gesellsschaftstrasse 6, 3012 Bern, Switzerland \\
$^{11}$ Instituto de Astrofisica de Canarias, 38200 La Laguna, Tenerife, Spain \\
$^{12}$ Departamento de Astrofisica, Universidad de La Laguna, 38206 La Laguna, Tenerife, Spain \\
$^{13}$ Institut de Ciencies de l'Espai (ICE, CSIC), Campus UAB, Can Magrans s/n, 08193 Bellaterra, Spain \\
$^{14}$ Institut d'Estudis Espacials de Catalunya (IEEC), 08034 Barcelona, Spain \\
$^{15}$ ESTEC, European Space Agency, 2201AZ, Noordwijk, NL \\
$^{16}$ Admatis, 5. Kandó Kálmán Street, 3534 Miskolc, Hungary \\
$^{17}$ Depto. de Astrofisica, Centro de Astrobiologia (CSIC-INTA), ESAC campus, 28692 Villanueva de la Cañada (Madrid), Spain \\
$^{18}$ Departamento de Fisica e Astronomia, Faculdade de Ciencias, Universidade do Porto, Rua do Campo Alegre, 4169-007 Porto, Portugal \\
$^{19}$ Space Research Institute, Austrian Academy of Sciences, Schmiedlstrasse 6, A-8042 Graz, Austria \\
$^{20}$ Center for Space and Habitability, Gesellsschaftstrasse 6, 3012 Bern, Switzerland \\
$^{21}$ INAF, Osservatorio Astronomico di Padova, Vicolo dell'Osservatorio 5, 35122 Padova, Italy \\
$^{22}$ Université Grenoble Alpes, CNRS, IPAG, 38000 Grenoble, France \\
$^{23}$ Institute of Planetary Research, German Aerospace Center (DLR), Rutherfordstrasse 2, 12489 Berlin, Germany \\
$^{24}$ Université de Paris, Institut de physique du globe de Paris, CNRS, F-75005 Paris, France \\
$^{25}$ Centre for Exoplanet Science, SUPA School of Physics and Astronomy, University of St Andrews, \\ North Haugh, St Andrews KY16 9SS, UK \\
$^{26}$ Centre for Mathematical Sciences, Lund University, Box 118, 221 00 Lund, Sweden \\
$^{27}$ Astrobiology Research Unit, Université de Liège, Allée du 6 Août 19C, B-4000 Liège, Belgium \\
$^{28}$ Space sciences, Technologies and Astrophysics Research (STAR) Institute, Université de Liège, Allée du 6 Août 19C, 4000 Liège, Belgium \\
$^{29}$ Leiden Observatory, University of Leiden, PO Box 9513, 2300 RA Leiden, The Netherlands \\
$^{30}$ Dipartimento di Fisica, Universita degli Studi di Torino, via Pietro Giuria 1, I-10125, Torino, Italy \\
$^{31}$ University of Vienna, Department of Astrophysics, Türkenschanzstrasse 17, 1180 Vienna, Austria \\
$^{32}$ Institut d'astrophysique de Paris, UMR7095 CNRS, Université Pierre \& Marie Curie, 98bis blvd. Arago, 75014 Paris, France \\
$^{33}$ Science and Operations Department - Science Division (SCI-SC), Directorate of Science, European Space Agency (ESA), \\ European Space Research and Technology Centre (ESTEC),
Keplerlaan 1, 2201-AZ Noordwijk, The Netherlands \\
$^{34}$ Konkoly Observatory, Research Centre for Astronomy and Earth Sciences, 1121 Budapest, Konkoly Thege Miklós út 15-17, Hungary \\
$^{35}$ ELTE E\"otv\"os Lor\'and University, Institute of Physics, P\'azm\'any P\'eter s\'et\'any 1/A, 1117 \\
$^{36}$ IMCCE, UMR8028 CNRS, Observatoire de Paris, PSL Univ., Sorbonne Univ., 77 av. Denfert-Rochereau, 75014 Paris, France \\
%$^{37}$ Department of Astrophysics, University of Vienna, Tuerkenschanzstrasse 17, 1180 Vienna, Austria \\
$^{38}$ INAF, Osservatorio Astrofisico di Catania, Via S. Sofia 78, 95123 Catania, Italy \\
$^{39}$ Institute of Optical Sensor Systems, German Aerospace Center (DLR), Rutherfordstrasse 2, 12489 Berlin, Germany \\
$^{40}$ Dipartimento di Fisica e Astronomia "Galileo Galilei", Universita degli Studi di Padova, Vicolo dell'Osservatorio 3, 35122 Padova, Italy \\
$^{41}$ ETH Zurich, Department of Physics, Wolfgang-Pauli-Strasse 2, CH-8093 Zurich, Switzerland \\
$^{42}$ Cavendish Laboratory, JJ Thomson Avenue, Cambridge CB3 0HE, UK \\
$^{43}$ Center for Astronomy and Astrophysics, Technical University Berlin, Hardenberstrasse 36, 10623 Berlin, Germany \\
$^{44}$ Institut für Geologische Wissenschaften, Freie Universit\"at Berlin, 12249 Berlin, Germany \\
$^{45}$ ELTE E\"otv\"os Lor\'and University, Gothard Astrophysical Observatory, 9700 Szombathely, Szent Imre h. u. 112, Hungary \\
$^{46}$ MTA-ELTE Exoplanet Research Group, 9700 Szombathely, Szent Imre h. u. 112, Hungary \\
$^{47}$ Institute of Astronomy, University of Cambridge, Madingley Road, Cambridge, CB3 0HA, United Kingdom
\\
$^{48}$ Lowell Observatory, 1400 W. Mars Hill Rd., Flagstaff, AZ 86001, USA \\

%%%%%%%%%%%%%%%%%%%%%%%%%%%%%%%%%%%%%%%%%%%%%%%%%%
% Don't change these lines
\bsp	% typesetting comment
\label{lastpage}

\end{document}